\documentclass[12pt]{article}
\usepackage[a4paper, margin=1in]{geometry}
\usepackage{amsmath, amssymb}
\usepackage{authblk}
\usepackage{graphicx}
\usepackage{xcolor}
\usepackage{cite}
\usepackage{hyperref}
\usepackage{subcaption}
\usepackage{float}
\usepackage[section]{placeins}
\usepackage{epstopdf}

\title{\textbf{Reconstruction of $f(G)$ Gravity from an Interacting Viscous Generalized QCD Ghost Dark Energy Model: Cosmology and Thermodynamics}}
\author[1]{Zhanar Umurzakhova\thanks{zhumurzakhova@gmail.com
}}
\author[2]{Aziza Altaibayeva\thanks{aziza.ltaibayeva@gmail.com 
}}
\author[3]{Ulbossyn Ualikhanova\thanks{ulbossyn.ualikhanova@gmail.com 
}}

\affil[1,2,3]{\small Department of General and Theoretical Physics, L.N. Gumilyov Eurasian National University, Astana 010008,Kazakhstan.}
\affil[1]{\small Mukhtar Auezov South Kazakhstan University, Tauke Khan Av, Shymkent, Kazakhstan}
\author[3]{Surajit Chattopadhyay \footnote{Corresponding author}\thanks{schattopadhyay1@kol.amity.edu; surajitchatto@outlook.com
}}
\affil[3]{Department of Mathematics, Amity University Kolkata, Major Arterial Road, Action Area II, Rajarhat, Newtown, Kolkata 700135, India }
\date{\today}

\begin{document}

\maketitle
\begin{abstract}
In this work, we investigate an interacting viscous generalized QCD ghost dark energy model in the framework of reconstructed $f(G)$ gravity proposed in \textit{Phys. Lett. B} \textbf{631}, 1--6 (2005). The interaction between dark matter and dark energy together with bulk viscosity is incorporated to describe a more realistic cosmic evolution. A hybrid expansion law is adopted to reconstruct the modified Gauss--Bonnet function, which naturally connects the early matter-dominated epoch with the present accelerated expansion of the universe. Since an exact analytical reconstruction is difficult, the $f(G)$ function is obtained numerically in both the early- and late-time regimes. Motivated by the numerical reconstruction, a reconstruction-inspired power-law form of $f(G)$ is also considered to examine the cosmological implications of the model. The results show that the reconstructed $f(G)$ function evolves smoothly throughout the cosmic history, while the effective equation of state gradually approaches the de Sitter phase at late times. The thermodynamic behaviour of the model is further examined using Barrow entropy following \textit{Eur. Phys. J. C} \textbf{81}, 644 (2021). The non-negative evolution of the total entropy shows the validity of the generalized second law of thermodynamics. The study finally concludes that the interacting viscous generalized QCD ghost dark energy model in reconstructed $f(G)$ gravity provides a viable and thermodynamically consistent framework for explaining the late-time accelerated expansion of the universe.
\end{abstract}

\tableofcontents

\section{Introduction}

The discovery of the accelerated expansion of the universe in 1998 from the observations of distant Type Ia supernovae (SNe) \cite{Riess1998,Perlmutter1999} changed our understanding of cosmology. Later, this result was supported by observations of the cosmic microwave background (CMB) and other cosmological data. The study of dynamical dark energy has received renewed attention in recent years due to the latest cosmological observations. The present accelerated expansion of the universe can be explained in two ways. One approach introduces dark energy with negative pressure within the framework of General Relativity (GR) \cite{Copeland2006,Bamba2012}. The other approach modifies General Relativity or develops alternative theories of gravity to explain the observed acceleration \cite{Clifton2012}. Abdalla \textit{et al.} \cite{Abdalla2022} presented a comprehensive review of the present cosmological tensions and anomalies, and discussed several theoretical models proposed to explain them. In the context of dark energy and modified gravity, Bhattacharjee \textit{et al. }\cite{Bhattacharjee2026} investigated the background cosmological evolution in a homogeneous matter creation scenario using the recent DESI DR2 BAO observations, where dark matter is produced through an adiabatic matter creation process as an alternative to dark energy and modified gravity. Recently, Sahoo \textit{et al.}~\cite{Sahoo2025} investigated a multi-component dark energy model consisting of a quintessence scalar field and a cosmological constant, and showed that the interaction between different dark energy components can significantly influence the cosmological dynamics. Other recent studies have shown that the interaction between dark matter and dark energy can significantly influence the cosmological evolution of the universe. In particular, Pradhan \textit{et al.}~\cite{Pradhan2025} investigated an interacting dark sector model in a curved FLRW scalar field cosmology and discussed its implications for the late-time dynamics of the universe. Recently, Duchaniya \textit{et al.}~\cite{Duchaniya2025} investigated interacting $\alpha$-attractor dark energy models and showed that dark-sector interactions can simultaneously account for the late-time accelerated expansion of the universe and influence the growth of large-scale structures. Agrawal \textit{et al.}~\cite{Agrawal2022} investigated a bouncing cosmological model in extended gravity and reconstructed it as an effective dark energy model, showing that modified gravity can successfully describe different phases of the cosmic evolution.

Cai \textit{et al.} \cite{Cai2026} reviewed different quintom dark energy models and discussed how the recent DESI DR2 observations favour an evolving dark energy equation of state. The cosmological constant was introduced as the simplest form of dark energy to explain this acceleration, leading to the standard $\Lambda$CDM model. Bamba \textit{et al.} \cite{Bamba2012} reviewed various dark energy cosmological models and discussed their physical properties as well as the observational tests used to explain the present accelerated expansion of the universe. Copeland \textit{et al.} \cite{Copeland2006} reviewed different dark energy models and their role in explaining the present accelerated expansion of the universe. In a recent study, de Cruz Pérez \textit{et al.} \cite{deCruzPerez2026} studied different dynamical dark energy models using the latest observational data and showed that some of these models perform better than the standard $\Lambda$CDM model.

In the literature of modified gravity, the Gauss--Bonnet (GB) invariant \cite{Koivisto2007, Nojiri2005} has become an important geometrical quantity. Due to its rich geometrical properties, the GB term has been widely used to study different stages of the evolution of the universe, including the early inflationary phase and the present accelerated expansion. It also appears naturally in the low-energy effective action of string-inspired gravity theories, making it an attractive framework for studying cosmological models. Modified Gauss--Bonnet gravity, or $f(G)$ gravity, was originally proposed as a viable modification of General Relativity to explain the late-time acceleration of the Universe without introducing an explicit dark energy component \cite{NojiriOdintsov_fG_2005,Cognola_fG_2006}. Since then, $f(G)$ gravity has attracted considerable attention in cosmology owing to its rich phenomenology, including inflationary dynamics, late-time cosmic acceleration, dark energy models, and thermodynamic aspects. A comprehensive review of the theoretical foundations and cosmological applications of modified gravity theories, including $f(G)$ gravity, can be found in Ref.~\cite{Nojiri_ModGrav_Review_2017}. Scalar--Gauss--Bonnet gravity has been shown to provide a viable inflationary scenario that agrees well with the latest observational constraints \cite{Odintsov2018}. To overcome the ghost instability problem, ghost-free formulations of modified Gauss--Bonnet gravity have also been proposed, making the theory theoretically consistent and physically viable \cite{Nojiri2019}. Moreover, the Gauss--Bonnet invariant has also been used to develop dark energy models motivated by string/M-theory, where the scalar field coupled with the GB term can explain the present accelerated expansion of the universe \cite{Nojiri2005}. Besides dark energy, modified Gauss--Bonnet gravity has also been used to study several important cosmological problems, including gravitational baryogenesis \cite{Odintsov2016}, mimetic gravity with Lagrange multiplier constraints \cite{Astashenok2015}, and inflationary models satisfying the swampland criteria while remaining compatible with the gravitational-wave event GW170817 \cite{Odintsov2023}. Li \textit{et al.} \cite{Li2007} investigated modified Gauss--Bonnet gravity and showed that the Gauss--Bonnet correction can explain the late-time acceleration of the universe. They \cite{Li2007} also pointed out that viable $f(G)$ models must satisfy the observational and stability conditions. Several studies have investigated the cosmological viability of modified $f(G)$ gravity. De Felice and Tsujikawa \cite{DeFelice2009} derived the conditions required for constructing cosmologically viable $f(G)$ gravity models and showed that the functional form of $f(G)$ should satisfy stability as well as observational constraints. The applications of $f(G)$ gravity have also been studied in different areas, such as cylindrical solutions \cite{Houndjo2014}, the stability of self-gravitating systems \cite{Yousaf2023}, and the energy conditions \cite{Garcia2011}. These studies show that $f(G)$ gravity is an important theory for studying different problems in cosmology and astrophysics. 

Till now ghost dark energy has been studied in many gravity models but viscous interacting ghost inside $f(G)$ gravity is not properly explored. Although the present work is motivated by the reconstruction approach adopted in ghost dark energy models within $f(G)$ gravity \cite{Sharif2019}, it differs from the previous study in several important aspects. The earlier work considered the standard ghost dark energy model with power-law form of scale factor and investigated the equation of state, classical stability, and statefinder diagnostics for interacting and non-interacting scenarios. In contrast, the present work reconstructs $f(G)$ gravity using an interacting viscous generalized QCD ghost dark energy model together with a hybrid expansion law that naturally describes the transition from the matter-dominated era to the late-time accelerated universe. Furthermore, we derive the hybrid expansion from the asymptotic behaviour of the model, perform separate numerical reconstructions of $f(G)$ in the early- and late-time regimes, and examine the cosmological evolution through the effective equation of state, stability analysis, and statefinder diagnostic. The present study also examines the thermodynamic consistency of the reconstructed model. These features mentioned above extend the earlier reconstruction approach \cite{Sharif2019} and provide a more detailed study of ghost dark energy in modified Gauss--Bonnet gravity.

In this work we try to study interacting viscous QCD ghost dark energy inside $f(G)$ gravity and test its stability and thermodynamical behaviour. To investigate the dynamics of the dark energy (DE) model, we consider a spatially flat Friedmann--Robertson--Walker (FRW) universe composed of three energy components, namely pressureless matter, dark energy, and radiation. In the ghost dark energy (GDE) framework, following reference \cite{Cai2012}, the energy density of the DE component is assumed to be $\rho_{\rm DE}=\alpha H+\beta H^{2},$ where $H$ is the Hubble parameter. The parameter $\alpha$ has the dimension of $[\mathrm{energy}]^{3}$ and is expected to be of the order of $\Lambda_{\rm QCD}^{3}$, where $\Lambda_{\rm QCD}\sim100\,\mathrm{MeV}$ denotes the Quantum Chromodynamics (QCD) mass scale \cite{Cai2012}. The second parameter, $\beta$, has the dimension of $[\mathrm{energy}]^{2}$ and represents the subleading contribution to the ghost dark energy density \cite{Cai2012}. The linear term in $H$ originates from the Veneziano ghost field in QCD and provides the dominant contribution at late cosmic times, whereas the quadratic term becomes significant during the early stages of cosmic evolution.

Although the present work focuses primarily on the late-time accelerated expansion of the Universe, modified $f(G)$ gravity has also been widely studied in the context of the early Universe, including inflationary cosmology and the generation of primordial perturbations \cite{Nojiri_ModGrav_Review_2017}. In particular, suitable forms of $f(G)$ gravity can successfully describe an inflationary phase and yield observational predictions for the scalar spectral index and the tensor-to-scalar ratio. However, the present reconstruction is based on a late-time interacting dark energy scenario, and therefore the analysis is restricted to the cosmological evolution, thermodynamic properties, and observational consistency of the reconstructed model. The rest of the paper is organized as follows. In Section~2, we have  presented the basic framework of $f(G)$ gravity. Section~3 introduces the viscous generalized QCD ghost dark energy model, while Section~4 discusses the interaction between dark matter and dark energy. The reconstruction procedure and the corresponding $f(G)$ function are developed in Sections~5 and~6, respectively. Sections~7 and~8 investigate the cosmological evolution and the classical stability of the reconstructed model. Section~9 examines the thermodynamic properties of the model and the validity of the generalized second law of thermodynamics. In Section~10, the theoretical predictions are compared with the 31 cosmic chronometer observations. Finally, Section~11 summarizes the main results and presents the concluding remarks.

\section{Basic Framework}

It has already been stated that the modified theories of gravity have become an important approach for explaining the accelerated expansion of the universe without introducing an unknown dark energy component \cite{Nojiri:2011,Capozziello:2011}. Among these theories, Gauss--Bonnet gravity has received considerable attention because it provides a natural extension of Einstein's General Relativity and can explain different stages of cosmic evolution \cite{Nojiri:2005,Fernandes2022}. The cosmological properties and observational constraints of Gauss--Bonnet gravity have been investigated in several works \cite{Amendola2006}, while the ghost-free conditions required for a physically acceptable Gauss--Bonnet cosmology have also been studied \cite{Calcagni2006}. Recently, modified gravity models at different curvature scales have been reviewed to understand their cosmological behaviour \cite{Mandal2025}. On the other hand, ghost dark energy and generalized ghost dark energy models have also been widely explored in different modified gravity theories \cite{Cai2012,Khurshudyan2015,Sharif2025}. These studies show that the combination of ghost dark energy and modified Gauss--Bonnet gravity provides a useful framework for studying the evolution of the universe. 

We consider flat FLRW universe which is homogeneous and isotropic in nature. This type of universe model is commonly used in cosmology to describe large scale behaviour of expanding universe \cite{Green2014,Holman2018,Koussour2022}. The line element for this spacetime is written as

\begin{equation}
ds^2=-dt^2+a^2(t)(dx^2+dy^2+dz^2)
\label{metric}
\end{equation}

where $a(t)$ is the scale factor which describes the expansion of the universe with time.

Making a proper quantum theory of gravity is very difficult work and till now only string theory (and also loop quantum gravity) can include gravity in quantum level in consistent way \cite{Calcagni2006}. So it becomes useful to study gravity at low energy level while also considering main corrections coming from string theory \cite{Calcagni2006}. This naturally takes us to study low-energy gravity actions which contain Gauss–Bonnet (GB) term, which is a special combination of squared curvature quantities from Riemann geometry \cite{Calcagni2006,Fernandes2022,Mandal2025}. In order to study the effect of modified gravity, we consider the action of $f(G)$ gravity given by

\begin{equation}
S=\int d^4x \sqrt{-g}
\left[
\frac{R}{2}+f(G)+\mathcal{L}_m
\right]
\label{action}
\end{equation}

Here $R$ is Ricci scalar and $f(G)$ is general function of Gauss--Bonnet invariant. $\mathcal{L}_m$ represents matter part of the universe.

For the above spacetime, the Gauss--Bonnet invariant takes the form

\begin{equation}
G=24H^2(\dot H+H^2)
\label{GB}
\end{equation}

where $H=\dot a/a$ is the Hubble parameter.

By varying the action given in Eq. (2) with respect to the metric, we obtain the explicit modified Friedmann equations for $f(G)$ gravity \cite{Nojiri:2005}:
\begin{equation}
3H^{2} = \rho_{m} + \rho_{ghost} + G f_{G} - f(G) - 24H^{3} \dot{f}_{G}
\label{field1}
\end{equation}
\begin{equation}
-2\dot{H} - 3H^{2} = p_{ghost} + f(G) - G f_{G} + 8H^{2} \ddot{f}_{G} + 16H \dot{H} \dot{f}_{G} + 16H^{3} \dot{f}_{G}
\label{field2}
\end{equation}
where $f_{G} = \frac{df}{dG}$ and the overdot denotes a derivative with respect to cosmic time $t$. These equations can be written in an effective fluid form as discussed in the literature \cite{Nojiri:2011,Capozziello:2011}:
\begin{equation}
3H^{2} = \rho_{m} + \rho_{eff}
\label{Equation:6}
\end{equation}
\begin{equation}
-2\dot{H} - 3H^{2} = p_{eff}
\end{equation}
where $\rho_{eff}$ and $p_{eff}$ represent the effective energy density and pressure contributions arising from the modified gravity sector. From the above equations, these effective components are identified as:
\begin{equation}
\rho_{eff} = \rho_{ghost} + G f_{G} - f(G) - 24H^{3} \dot{f}_{G}
\label{8}
\end{equation}
\begin{equation}
p_{eff} = p_{ghost} + f(G) - G f_{G} + 8H^{2} \ddot{f}_{G} + 16H \dot{H} \dot{f}_{G} + 16H^{2} \dot{f}_{G}
\label{9}
\end{equation}
This mathematical structure provides the basis for the reconstruction of the $f(G)$ functional form, which will be achieved by equating the geometric contributions to the energy density of the ghost dark energy model in a later section.

\section{Viscous QCD Ghost Dark Energy}

It is found that the contribution of the Veneziano QCD ghost field to the vacuum energy is not exactly proportional to the Hubble parameter $H$. There may also appear a smaller correction term proportional to $H^2$. This extended form is usually known as the generalized ghost dark energy (GGDE) model. Therefore, the energy density of GGDE can be written as \cite{Cai2012, Sharif2025}

\begin{equation}
\rho_{ghost}=\alpha H+\beta H^2 ,
\label{Eq:6}
\end{equation}

where $H$ denotes the Hubble parameter, and $\alpha$ and $\beta$ are constant parameters characterizing the contribution of the ghost dark energy component. The first term, proportional to $H$, represents the leading contribution associated with the QCD ghost field in a dynamical background spacetime, while the second term proportional to $H^2$ is often considered as a sub-leading correction which may arise due to quantum effects or higher-order contributions \cite{Cai2012,Sharif2025,Khurshudyan2015}. It is important to note that the QCD ghost dark energy model considered in this work is different from the usual scalar field dark energy models. According to the original studies by Urban and Zhitnitsky \cite{Urban2009} and Cai \textit{et al.} \cite{Cai2012}, the QCD ghost field comes from the topological structure of the QCD vacuum. In flat Minkowski spacetime, this ghost field is not a physical field. It does not propagate or introduce any new physical particles. Instead, it remains decoupled from the physical spectrum and helps preserve the consistency of the quantum field theory.

In order to incorporate dissipative effects in the cosmic fluid, we include the effect of bulk viscosity. In the presence of bulk viscosity, the effective pressure of the ghost dark energy component is modified and can be written as

\begin{equation}
p_{ghost}=w_{ghost}\rho_{ghost}-3H\xi ,
\label{Eq:7}
\end{equation}

where $w_{ghost}$ is the equation of state parameter of the ghost dark energy and $\xi$ represents the bulk viscosity coefficient. The additional term $-3H\xi$ accounts for the viscous pressure arising due to irreversible processes in the cosmic fluid.

Following a commonly used phenomenological assumption \cite{Brevik2017,Yang2022,Singh2018,Brevik2022}, the bulk viscosity coefficient is taken as a linear function of the Hubble parameter, namely
\begin{equation}
\xi=\xi_0+\xi_1 H,
\label{Eq:8}
\end{equation}
where $\xi_0$ and $\xi_1$ are constant parameters. The parameter $\xi_0$ corresponds to a constant viscosity contribution, whereas the term proportional to $H$ represents the possible dependence of the viscous effects on the cosmic expansion rate. To explain why we have chosen this form of $\xi$, we refer to the review by Brevik et al. \cite{Brevik2017}, where the general form was considered as $\xi = \xi_{0} + \xi_{1} H + \xi_{2}\frac{\dot{H}}{H}$, and \eqref{Eq:8} is a particular case of it.

Such a parametrization of the bulk viscosity has been widely used in cosmological studies, since it allows the viscous pressure to evolve with the expansion of the universe and provides a simple but effective way to investigate dissipative effects in dark energy models.

\section{Interacting Viscous Generalized Ghost Dark Energy Model}
In the present framework, we assume that dark matter and generalized ghost dark energy interact with each other through a phenomenological coupling term. Such interacting dark energy models have received considerable attention \cite{He2008,Wang2016,JSWang2014} in cosmology since they may help in explaining the late-time accelerated expansion of the universe and can also provide a possible resolution to the cosmic coincidence problem \cite{Hu2006,Sadjadi2006,Caldera2009,Berger2006}. In addition, we incorporate the effect of bulk viscosity in the dark energy sector in order to account for dissipative processes in the cosmic fluid \cite{Avelino2025}. Under these assumptions, the conservation equations for dark matter and ghost dark energy are written as

\begin{equation}
\dot\rho_m+3H\rho_m=Q
\label{Eq:9}
\end{equation}

\begin{equation}
\dot\rho_{ghost}+3H(1+w_{ghost})\rho_{ghost}
=-Q+9H^2\xi
\label{Eq:10}
\end{equation}

Equation~(\ref{Eq:9}) describes the evolution of dark matter in the presence of interaction, while Eq.~(\ref{Eq:10}) governs the dynamics of ghost dark energy including both interaction and bulk viscous effects. The additional term $9H^2\xi$ arises due to the presence of bulk viscosity and represents dissipative effects in the cosmic fluid. We choose the interaction term $Q$ in the form (for details see \cite{He2008} and references therein)

\begin{equation}
Q=3b^2H(\rho_m+\rho_{ghost})
\label{Eq:11}
\end{equation}

where $b^2$ is the coupling parameter that determines the strength of interaction between dark matter and ghost dark energy. The interaction term $Q$ represents the rate of energy exchange between the dark matter and dark energy sectors \cite{Mishra2023}. A positive value of $Q$ ($Q>0$) indicates that energy is transferred from dark matter to dark energy, whereas a negative value ($Q<0$) corresponds to energy transfer in the opposite direction \cite{Mishra2023}. Although the interaction term determines the energy exchange between the two dark sectors, its exact physical origin is still unknown. Therefore, phenomenological forms of $Q$ are commonly adopted to study the possible effects of dark-sector interaction on the cosmological evolution. The chosen interaction term is phenomenological in nature and has been widely used in the literature because it leads to stable cosmological evolution and allows a possible resolution of the cosmic coincidence problem.

Using Eq.~(\ref{Eq:6}), the time derivative of the ghost energy density is obtained as
\begin{equation}
\dot{\rho}_{ghost} = (\alpha + 2\beta H)\dot{H}.
\label{Eq:12}
\end{equation}

Substituting Eqs.~(\ref{Eq:6}) and (\ref{Eq:8}) into the conservation equation (\ref{Eq:10}), we obtain
\begin{equation}
(\alpha + 2\beta H)\dot{H} + 3H(1+w_{ghost})(\alpha H + \beta H^2)
= -Q + 9H^2(\xi_0 + \xi_1 H).
\label{Eq:13}
\end{equation}

Now, using the interaction term from Eq.~(\ref{Eq:11}), Eq.~(\ref{Eq:13}) can be written as
\begin{equation}
(\alpha + 2\beta H)\dot{H} + 3H(1+w_{ghost})(\alpha H + \beta H^2)
= -3b^2H(\rho_m+\rho_{ghost}) + 9H^2(\xi_0 + \xi_1 H).
\label{Eq:14}
\end{equation}

Solving Eq.~(\ref{Eq:14}) for the equation of state parameter $w_{ghost}$, we obtain
\begin{equation}
w_{ghost} =
-1 + \frac{-3b^2H(\rho_m+\rho_{ghost})
+ 9H^2(\xi_0 + \xi_1 H)
- (\alpha + 2\beta H)\dot{H}}
{3H(\alpha H + \beta H^2)}.
\label{Eq:15}
\end{equation}

Further, substituting $\rho_{ghost}$ from Eq.~(\ref{Eq:6}), we obtain the explicit form
\begin{equation}
w_{ghost} =
-1 + \frac{-3b^2H\left(\rho_m+\alpha H + \beta H^2\right)
+ 9H^2(\xi_0 + \xi_1 H)
- (\alpha + 2\beta H)\dot{H}}
{3H(\alpha H + \beta H^2)}.
\label{Eq:16}
\end{equation}

It is important to note that the matter density $\rho_m$ appearing in Eq.~(\ref{Eq:16}) is not independent, but evolves according to the interaction equation (\ref{Eq:9}). Therefore, $\rho_m$ must be determined consistently by solving Eq.~(\ref{Eq:9}) with the chosen interaction term (\ref{Eq:11}), rather than assuming the standard scaling $\rho_m \propto a^{-3}$. This ensures that the effects of interaction between dark matter and ghost dark energy are fully incorporated into the cosmological dynamics. From the expression of Eq.~\eqref{Eq:16}, we further see that the equation of state parameter of ghost dark energy is influenced by interaction between dark matter and dark energy, viscous effects, and the dynamical evolution of the Hubble parameter.
Now, Using the interaction term given in Eq.~(\ref{Eq:11}), the conservation equation for dark matter (Eq.~(\ref{Eq:9})) becomes
\begin{equation}
\dot{\rho}_m + 3H\rho_m = 3b^2 H(\rho_m + \rho_{ghost}).
\label{Eq:17}
\end{equation}

Rewriting Eq.~(\ref{Eq:17}), we obtain
\begin{equation}
\dot{\rho}_m + 3H(1 - b^2)\rho_m = 3b^2 H \rho_{ghost}.
\label{Eq:18}
\end{equation}

Substituting $\rho_{ghost}$ from Eq.~(\ref{Eq:6}), Eq.~(\ref{Eq:18}) takes the form
\begin{equation}
\dot{\rho}_m + 3H(1 - b^2)\rho_m = 3b^2 H(\alpha H + \beta H^2).
\label{Eq:19}
\end{equation}

Equation~(\ref{Eq:19}) is a first-order linear differential equation for $\rho_m$. 
Using the integrating factor method, the integrating factor is given by
\begin{equation}
\mu(t) = \exp\left[\int 3H(1-b^2)\,dt\right].
\label{Eq:20}
\end{equation}

Thus, the solution for $\rho_m$ can be written as
\begin{equation}
\rho_m = \mu^{-1}(t) \left[ \int 3b^2 H(\alpha H + \beta H^2)\mu(t)\,dt + C \right],
\label{Eq:21}
\end{equation}
where $C$ is the constant of integration. Eq.~\eqref{Eq:21} shows that dark matter does not evolve as $\rho_m \propto a^{-3}$ in the present model. Its evolution is affected by the interaction with ghost dark energy as already mentioned and also by the expansion of the universe. Due to the presence of the interaction term, energy transfer takes place between dark matter and dark energy, so the matter density depends on the past evolution of the universe. The equation also shows that the interaction strength, Hubble expansion, and initial conditions together control the behaviour of matter density during cosmic evolution.

\section{Reconstruction}

In this section, we consider the interacting viscous ghost dark energy model in $f(G)$ gravity to reconstruct our scale factor $a(t)$. The aim is to derive the functional form of the scale factor from the underlying evolution equations in the early- and late-time limits, instead of assuming any particular phenomenological form of it. In the early universe, we take the limit $t \to 0$, and here the Hubble parameter $H \gg 1$ and in such scenario, the ghost dark energy density $\rho_{\text{ghost}} = \alpha H + \beta H^2$ is dominated by the quadratic term, i.e., $\rho_{\text{ghost}} \simeq \beta H^2$ as $H$ becomes less significant compared to $H^2$. The $w_{\text{eff}}$ can thus be written in the standard form
\begin{equation}
w_{\text{eff}} = -1 - \frac{2\dot{H}}{3H^2}.
\end{equation}
To keep consistency with standard cosmology, the model should show a matter-dominated phase in the early universe. This means that at early time, the effective equation of state should go close to $w_{\text{eff}} \approx 0$. From this we have  $\dot{H} \simeq -\frac{3}{2}H^2$, which on integration produces $H(t) = \frac{2}{3t}$. Using the relation $H = \dot{a}/a$, the corresponding scale factor is found to be $a(t) \propto t^{2/3}$, which reproduces the standard matter-dominated expansion and shows that the model reproduces the early-time limit. In the opposite limit $t \to \infty$, that is at late time, the Hubble parameter changes very slowly and becomes almost constant. In this case, the linear term in $\rho_{\text{ghost}}$ becomes dominant and the term involving $\dot{H}$ can be neglected. As a result, the effective equation of state goes to $w_{\text{eff}} \to -1$, which shows a de Sitter type phase. At late time, we can take $\dot{H} \to 0$, which means the Hubble parameter becomes nearly constant, i.e., $H \to \lambda$. Then from the relation $\frac{\dot{a}}{a} = \lambda$, we get the solution $a(t) \propto e^{\lambda t}$. This shows the accelerated expansion of the universe at late time due to dark energy. Therefore, we can say that the cosmological behaviour of the present model changes smoothly from a power-law expansion $a(t) \sim t^{2/3}$ at early time to an exponential expansion $a(t) \sim e^{\lambda t}$ at late time. Thus, combining the two regimes, the functional form that can reproduce both limits is 
\begin{equation}
a(t) = a_0 t^m e^{\lambda t}.
\label{Eq:22}
\end{equation}
Hence, within the present interacting viscous $f(G)$ framework under consideration, the hybrid scale factor is chosen for further analysis. Then

\begin{equation}
H=\frac{m}{t}+\lambda
\label{Eq:23}
\end{equation}

\begin{equation}
\dot H=-\frac{m}{t^2}
\label{Eq:24}
\end{equation}

Substituting Eqs.~(\ref{Eq:23}) and (\ref{Eq:24}) into Eq.~(\ref{Eq:16}), the equation of state parameter can be reconstructed explicitly in terms of cosmic time.

Using
\[
H=\frac{m}{t}+\lambda, \qquad \dot{H}=-\frac{m}{t^2},
\]
we obtain

\begin{equation}
w_{ghost}(t)=-1 + \frac{-3b^2 H\left(\rho_m+\alpha H + \beta H^2\right)+ 9H^2(\xi_0 + \xi_1 H)+ \frac{m}{t^2}(\alpha + 2\beta H)}{3H(\alpha H + \beta H^2)},
\label{Eq:25}
\end{equation}

where $H=\frac{m}{t}+\lambda$.

Further, substituting the reconstructed matter density $\rho_m$ from Eq.~(\ref{Eq:21}), the above expression becomes

\begin{equation}
\begin{aligned}
w_{ghost}(t) = -1 + \frac{1}{3H(\alpha H + \beta H^2)} \Bigg[& -3b^2 H \Bigg\{ \mu^{-1}(t)\left( \int 3b^2 H(\alpha H + \beta H^2)\mu(t)\,dt + C \right) \\
& \quad + \alpha H + \beta H^2 \Bigg\} \\
& + 9H^2(\xi_0 + \xi_1 H) + \frac{m}{t^2}(\alpha + 2\beta H)
\Bigg].
\end{aligned}
\label{Eq:26}
\end{equation}
Here, the integrating factor $\mu(t)$ is given by Eq.~(\ref{Eq:20}), which under the hybrid scale factor takes the explicit form
\begin{equation}
\mu(t)=\exp\left[3(1-b^2)\int \left(\frac{m}{t}+\lambda\right)dt\right]
= t^{3m(1-b^2)} e^{3\lambda(1-b^2)t}.
\label{Eq:27}
\end{equation}
Substituting the hybrid scale factor given in Eq.~(\ref{Eq:22}) into the Hubble parameter and its derivative, we have
\begin{equation}
H=\frac{m}{t}+\lambda, \qquad \dot{H}=-\frac{m}{t^2}.
\end{equation}
Using these relations in Eq.~(\ref{Eq:26}), the equation of state parameter can be expressed entirely as a function of cosmic time. After straightforward substitution, we obtain
\begin{equation}
\begin{aligned}
w_{\text{ghost}}(t) = -1 &+ \frac{1}{3 \left( \frac{m}{t} + \lambda \right) \left[ \alpha \left( \frac{m}{t} + \lambda \right) + \beta \left( \frac{m}{t} + \lambda \right)^2 \right]} \Biggl[ -3b^2 \left( \frac{m}{t} + \lambda \right) \\
& \times \left\{ \mu^{-1}(t) \left( \int 3b^2 \left( \frac{m}{t} + \lambda \right) \left[ \alpha \left( \frac{m}{t} + \lambda \right) + \beta \left( \frac{m}{t} + \lambda \right)^2 \right] \mu(t) \, dt + C \right) \right. \\
& \left. + \alpha \left( \frac{m}{t} + \lambda \right) + \beta \left( \frac{m}{t} + \lambda \right)^2 \right\} \\
& + 9 \left( \frac{m}{t} + \lambda \right)^2 \left( \xi_0 + \xi_1 \left( \frac{m}{t} + \lambda \right) \right) + \frac{m}{t^2} \left[ \alpha + 2\beta \left( \frac{m}{t} + \lambda \right) \right] \Biggr].
\end{aligned}
\label{Eq:28}
\end{equation}

It may be noted that $\mu(t)$ in \eqref{Eq:28} is same as given in \eqref{Eq:27}. Equation~(\ref{Eq:28}) represents the fully reconstructed equation of state parameter of the viscous interacting ghost dark energy model in terms of cosmic time. Eq. \eqref{Eq:28} clearly shows that the evolution of $w_{ghost}(t)$ is governed by the combined effects of the interaction between dark matter and dark energy through the coupling parameter $b^2$, the dissipative contributions arising from bulk viscosity via $\xi_0$ and $\xi_1$, and the dynamical behaviour of $H$ coming from the hybrid scale factor. This interplay is expected to allow the possibility of transition from quintessence to phantom regime depending on the choice of model parameters. Also, Eq.~(\ref{Eq:28}) gives the time-dependent equation of state parameter of the viscous interacting ghost dark energy model for the hybrid scale factor. From this expression, we can further see how the interaction between dark matter and dark energy and bulk viscosity (as represented through $\xi_0$, $\xi_1$) influence the evolution of the universe.

In the RHS of Eq.~\eqref{Eq:26}, the first significant contribution comes from the interaction term containing $b^2$. This term has an integral with the integrating factor $\mu(t)$, which shows how energy is exchanged between dark matter and ghost dark energy over time. From physical perspective, it means that the present value of the equation of state depends not only on the current values but also on the past evolution of the interaction. Due to this, the interaction can change the effective pressure of dark energy and may push the model towards phantom behaviour ($w_{\text{ghost}}<-1$) when the coupling is large. 
The second term of the RHS of Eq.~\eqref{Eq:26} i.e. $9H^2(\xi_0+\xi_1 H)$ comes due to bulk viscosity effect. This term produces extra negative pressure in the cosmic fluid because of dissipative nature. As it depends on higher power of Hubble parameter, its effect is more at early time when expansion is high. So, viscosity helps in increasing the acceleration of the universe and also can help the model to go into phantom region. The third term in Eq.~\eqref{Eq:26} i.e. $\frac{m}{t^2}(\alpha+2\beta H)$ comes due to change of $H$ with time. This term is very large at early time ($t \to 0$), so equation of state changes very fast in early universe. It shows that the hybrid scale factor is dynamical and it connects deceleration phase and acceleration phase smoothly. For the numerical analysis, representative values of the model parameters are chosen to study the behaviour of the reconstructed model. These values give physically reasonable results and are similar to those commonly used in the literature for interacting dark energy and viscous cosmological models. The main purpose of this work is to investigate the theoretical behaviour of the model rather than to determine the best-fit values of the parameters. The values of the parameters would be mentioned in the subsequent part of the paper.
\begin{figure}[ht]
\centering
\includegraphics[width=0.75\textwidth]{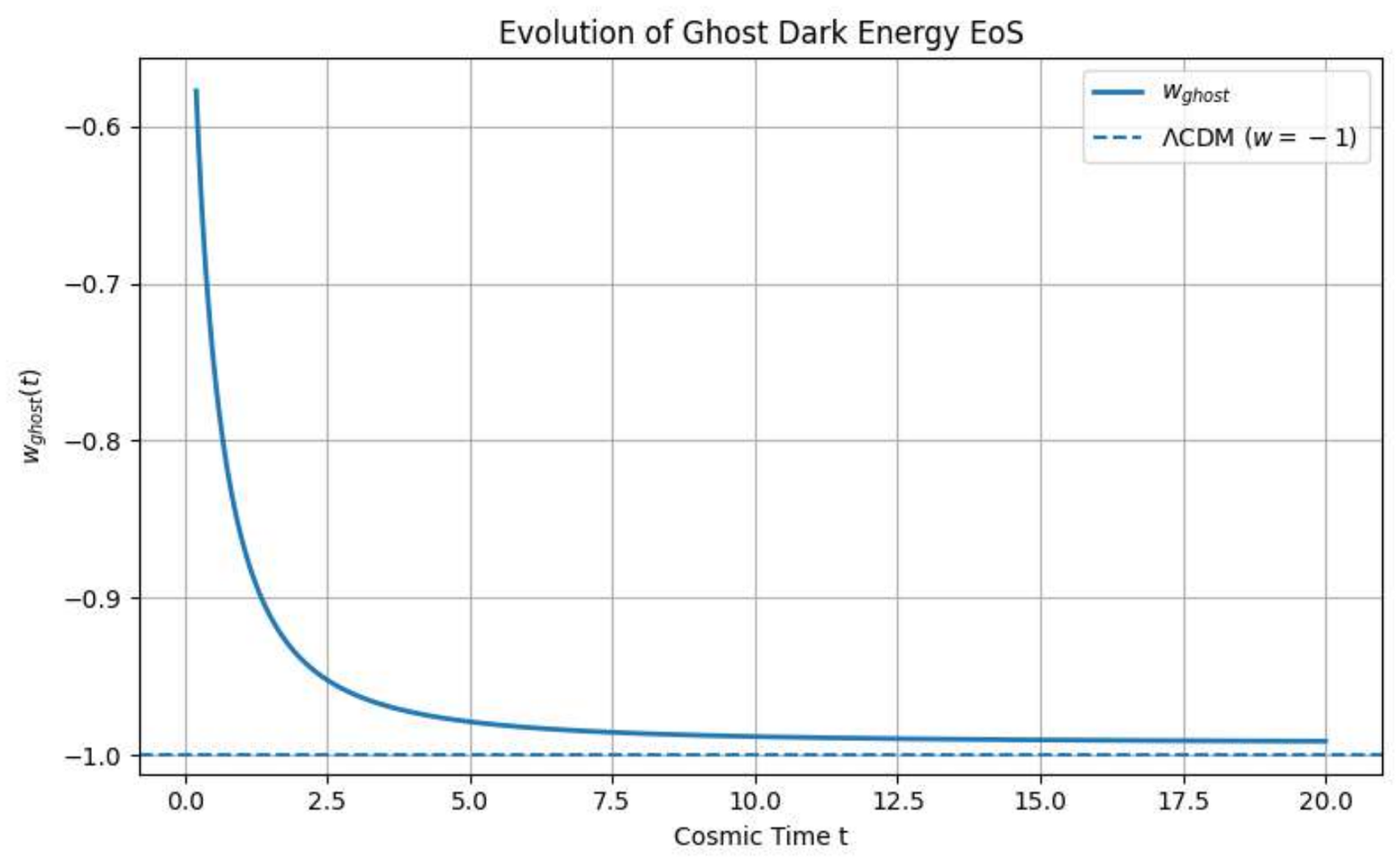}
\caption{
Evolution of the equation of state parameter $w_{\rm ghost}$ of the interacting viscous generalized QCD ghost dark energy model described by Eq.~\eqref{Eq:28} as a function of cosmic time $t$. The numerical analysis has been carried out for the parameter values $m=2/3$, $\lambda=0.75$, $\alpha=1.0$, $\beta=0.10$, $b^{2}=0.03$, $\xi_{0}=0.01$, $\xi_{1}=0.005$, and $C=0.5$. The horizontal dashed line corresponds to the $\Lambda$CDM limit $w=-1$. The evolution illustrates the combined influence of the dark-sector interaction, bulk viscous effects, and the hybrid expansion law on the dynamical behaviour of ghost dark energy. The asymptotic approach of $w_{\rm ghost}$ towards $-1$ at late times indicates convergence to a dark-energy-dominated accelerating epoch.}
\label{Fig:figure1}
\end{figure}

The Fig.~\ref{Fig:figure1} pictorially depicts the evolution of the equation of state parameter $w_{\rm ghost}$ of the interacting viscous generalized QCD ghost dark energy model as a function of cosmic time. At early times, $w_{\rm ghost}$ takes values greater than $-1$, indicating a quintessence-like behaviour of the dark energy component. From this figure it is observed that $w_{\rm ghost}$ has monotonic decreasing pattern that gradually approaches the $\Lambda$CDM value $w=-1$ and becomes asymptotic. This behaviour can be interpreted as a combined effect of the interaction between dark matter and dark energy, together with bulk viscosity that drives the cosmic evolution towards a dark-energy-dominated phase. The smooth evolution of $w_{\rm ghost}$ without any discontinuity or sudden abrupt changes also indicates that the reconstructed model remains physically amooth throughout the cosmic evolution.

\section{Reconstruction Scheme for $f(G)$ Gravity}

To reconstruct the functional form of $f(G)$ gravity that is compatible with the viscous interacting ghost dark energy model, we establish a correspondence between the geometric modification and the energy density of the dark sector. By comparing the modified Friedmann equations with the effective form as already mentioned, we identify the energy density contribution arising from the $f(G)$ sector as \cite{Nojiri:2011,Capozziello:2011}:
\begin{equation}
\rho_{eff} = \rho_{ghost} + G f_{G} - f(G) - 24H^{3} \dot{f}_{G} 
\label{Equation:35}
\end{equation}
The basic reconstruction is founded on the requirement that the late-time acceleration be driven by the combined effect of the modified gravity and the ghost field. Consequently, by equating the geometric terms to the generalized ghost dark energy density $\rho_{ghost}$ defined in Eq. \eqref{Eq:6}, we obtain the following master differential equation for the reconstruction:
\begin{equation}
f(G) - G f_{G} + 24H^{3} \dot{f}_{G} = \rho_{ghost} \quad 
\label{eq: 36}
\end{equation}
where $\rho_{ghost} = \alpha H + \beta H^{2}$ as already stated in previous section. From this equation we can proceed for determining specific functional form of $f(G)$ in the framework of hybrid scale factor. In this setup, we consider the modified gravity as a geometric equivalent to the ghost dark energy. As we are doing the same in the framework of hybrrid expansion law, we expect to have  a transition from the matter-dominated era to the current accelerated expansion under the cosmological settings of the reconstructed $f(G)$ with hybrid scale factor.

As the first step towards the reconstruction procedure, we first reconstruct the functional form of $f(G)$. In this phase, we will express the Gauss--Bonnet invariant in terms of the cosmic time $t$. For a spatially flat FLRW spacetime, the Gauss--Bonnet invariant can be expressed as a function of cosmic time $t$ as follows: 
\begin{equation}
G(t)=24\left(\frac{m}{t}+\lambda\right)^{2}\left[\left(\frac{m}{t}+\lambda\right)^{2}-\frac{m}{t^{2}}\right].
\label{Gt}
\end{equation}
The above Equation~(\ref{Gt}) gives the time-evolutionary behaviour of the Gauss--Bonnet invariant for the adopted hybrid expansion law. From  (\ref{Gt}) it follows that the curvature invariant is determined by the parameters $m$ and $\lambda$, which respectively govern the early-time power-law expansion and the late-time de Sitter-like evolution. We now discuss the asymptotic behaviour \cite{Kiyani2006} of $G(t)$. In the early universe ($t\rightarrow0$), the power-law contribution dominates over the
exponential term. This results in 
\begin{equation}
H\simeq\frac{m}{t},
\qquad
\dot H\simeq-\frac{m}{t^{2}},
\end{equation}
from which we can see that Eq.~(\ref{Gt}) reduces to 
\begin{equation}
G\simeq 24m^{3}(m-1)t^{-4}.
\label{Gearly}
\end{equation}
Thus, the Gauss--Bonnet invariant diverges as $t^{-4}$ during the early
matter-dominated epoch. From this we can interpret that higher-curvature corrections become significant at very early times. In the opposite limit, i.e. if we take ($t\rightarrow\infty$), the $H$ approaches the constant value as
\begin{equation}
H\rightarrow\lambda,
\qquad
\dot H\rightarrow0.
\end{equation}
Consequently Eq.~(\ref{Gt}) approaches 
\begin{equation}
G\rightarrow24\lambda^{4},
\label{Glate}
\end{equation}
which corresponds to a constant-curvature de Sitter spacetime. Therefore, in the chosen hybrid expansion law, we have an interpolation between a high-curvature early universe and a constant-curvature accelerating phase at late times. This makes the hybrid scale factor a suitable choice for reconstructing viable $f(G)$ gravity models capable of describing the complete cosmic evolution.

\subsection*{Early-Time Reconstruction}
We consider the case of $t\rightarrow0$. We have already stated that the exponential contribution in the hybrid scale factor becomes negligible  andthe Gauss--Bonnet invariant becomes $G \simeq 24m^{3}(m-1)t^{-4}$
as already shown. From this limitig approximation of $G$, we can write 
\begin{equation}
t=\left[\frac{24m^{3}(m-1)}{G}\right]^{1/4}.
\label{tinverse}
\end{equation}
Using Eq.~(\ref{tinverse}), we can have 
\begin{equation}
H(G)=m\left[\frac{G}{24m^{3}(m-1)}\right]^{1/4},
\end{equation}
Then we have the form of $\rho_{\rm ghost}(G)$ as 
\begin{equation}
\rho_{\rm ghost}(G)=\alpha m \left(\frac{G}{24m^{3}(m-1)}
\right)^{1/4}+\beta m^{2}\left(\frac{G}{24m^{3}(m-1)}
\right)^{1/2}.
\label{rhoG}
\end{equation}

The reconstruction equation therefore becomes a second-order differential equation for the unknown function $f(G)$:
\begin{equation}
f(G)-Gf_{G}+24H^{3}\dot G\,f_{GG}
=\rho_{\rm ghost}(G)
\label{master}
\end{equation}
This equation can not be solved analytically because of its highly nonlinear structure. However, from Eqs.~(\ref{rhoG})--(\ref{master})
we have mathematical framework required for numerical
reconstruction. From the above asymptotic analysis we can see the physical consistency of the reconstruction scheme and also we understand that the modified Gauss--Bonnet contribution is dominant during the high-curvature regime and asymptotically approaches to an effective cosmological constant in the late-time de Sitter phase. A complete numerical reconstruction of $f(G)$ is now carried out from Eq.~(\ref{master}) for some physically viable choice of the model parameters.

For convenience, first we write 
\begin{equation}
F(t)\equiv f(G(t)), \nonumber
\end{equation}
to avoid the explicit inversion of $G(t)$. Using the chain rule we have
\begin{equation}
f_{G}=\frac{\dot{F}}{\dot{G}},\qquad
f_{GG}=\frac{\ddot{F}\dot{G}-\dot{F}\ddot{G}}{\dot{G}^{3}},
\end{equation}
Eq.~(\ref{master}) is transformed into a second-order ordinary differential
equation for $F(t)$. The resulting equation is integrated numerically with suitable initial conditions, and the reconstructed functional form of $f(G)$ is obtained.
\begin{figure}[ht]
\centering

\begin{subfigure}{0.48\textwidth}
    \centering
    \includegraphics[width=\linewidth]{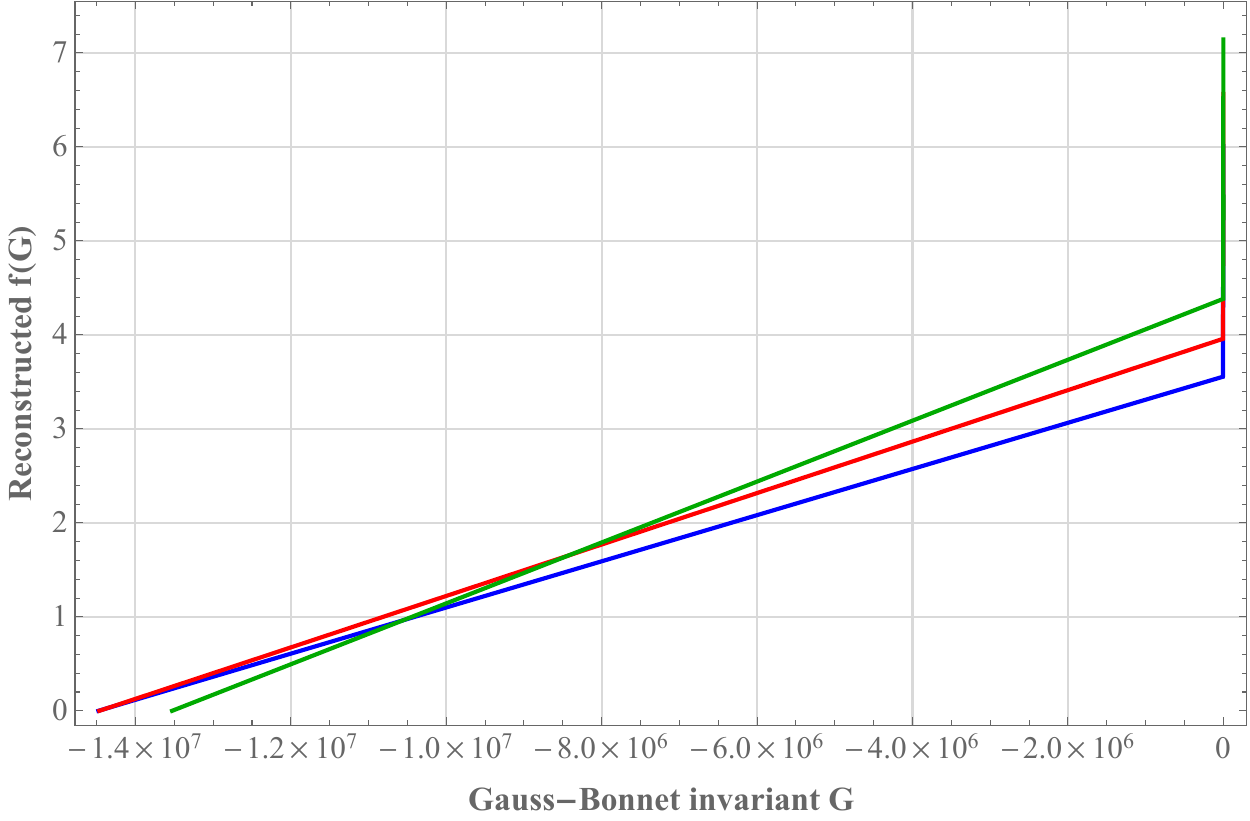}
    \caption{Early-time reconstruction.}
    \label{fig:Fig2a}
\end{subfigure}
\hfill
\begin{subfigure}{0.48\textwidth}
    \centering
    \includegraphics[width=\linewidth]{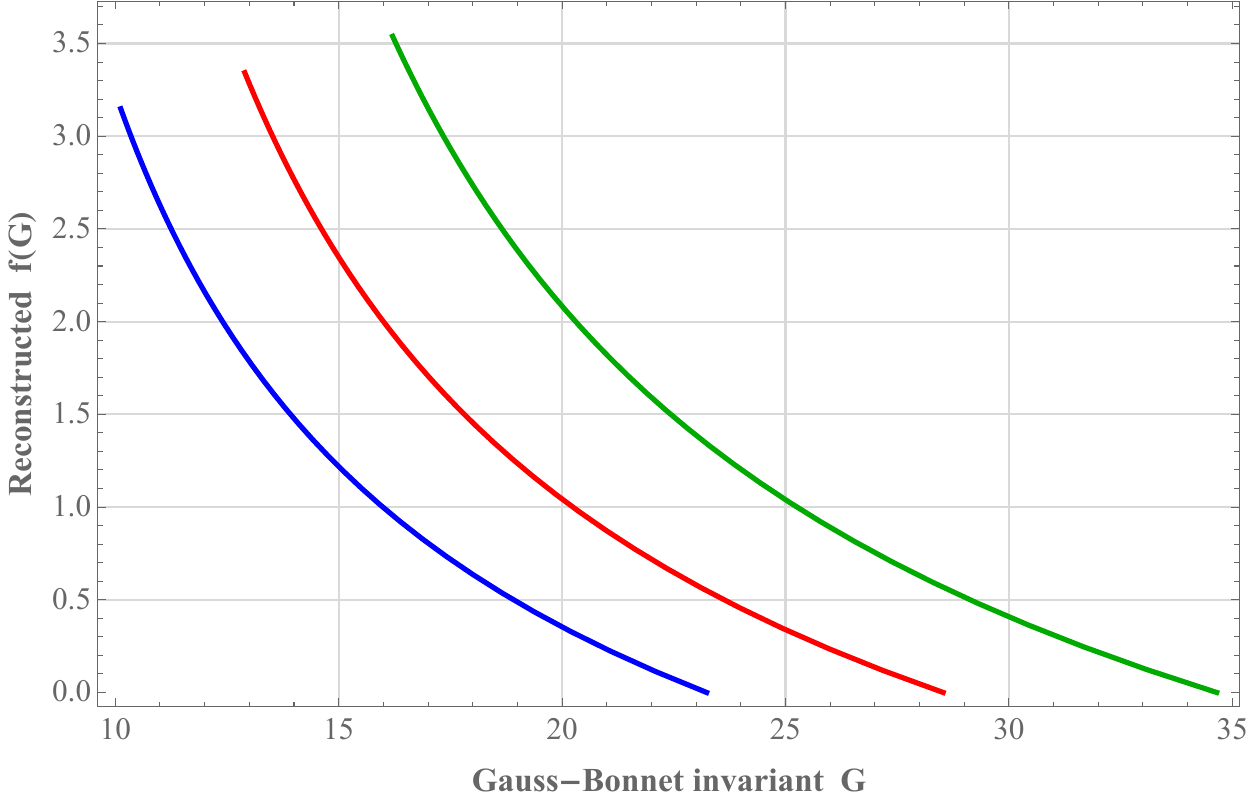}
    \caption{Late-time reconstruction.}
    \label{fig:Fig2b}
\end{subfigure}

\caption{Reconstructed modified Gauss--Bonnet gravitational function $f(G)$ versus the Gauss--Bonnet invariant $G$ obtained from the numerical reconstruction procedure for the hybrid scale factor. Panel (a) corresponds to the early-time epoch, where the curves are plotted for $m=0.70$ (blue), $m=0.75$ (red), and $m=0.80$ (green), with $\lambda=0.75$, $\alpha=1.0$, and $\beta=0.10$ fixed. Panel (b) corresponds to the late-time epoch, where the curves are plotted for $\lambda=0.70$ (blue), $\lambda=0.75$ (red), and $\lambda=0.80$ (green), with $m=0.67$, $\alpha=1.0$, and $\beta=0.10$ fixed. The comparison illustrates the sensitivity of the reconstructed function to variations in the dominant expansion parameter during the early- and late-time cosmological evolution.}
\label{fig:Fig2}
\end{figure}

The reconstructed $f(G)$ is pictorially presented in Fig.~\ref{fig:Fig2a} for early-time reconstruction.  Fig.~\ref{fig:Fig2} shows a smooth and monotonic behaviour of reconstructed $f(G)$ against $G$ throughout the early universe that corresponds to high-curvature regime. The reconstructed $f(G)$ is seen to increase almost linearly with $G$. This indicates that the modified Gauss--Bonnet contribution remains regular even in the presence of strong curvature effects. Furthermore, increasing the hybrid expansion parameter from $m=0.70$ to $m=0.80$ leads to a steeper reconstructed profile. Overall, we observe the reconstruction remains numerically stable and physically viable over the considered high-curvature domain. 

\subsubsection*{Late-Time Reconstruction}

We now consder the asymptotic behaviour of the reconstructed modified
Gauss--Bonnet gravity corresponding to the late-time limit ($t\rightarrow\infty$). In this regime, it has already been shown that the exponential component of the hybrid scale factor dominates the cosmic
evolution, and 
\[
H\simeq\lambda,\qquad \dot H\simeq0.
\]
Consequently, the Gauss--Bonnet invariant becomes
\[
G\simeq24\lambda^{4},
\]
Since the generalized ghost dark energy density is given by $\rho_{\rm ghost}=\alpha H+\beta H^{2}$, its late-time value also approaches a constant,
\[
\rho_{\rm ghost}\simeq\alpha\lambda+\beta\lambda^{2}.
\]

Under these asymptotic conditions, the reconstruction equation \eqref{eq: 36} reduces to a differential equation with nearly constant coefficients. Rather than seeking an analytical approximation, we solve the reconstruction equation numerically using
the same procedure adopted in the case of early-time reconstruction. Following the numerical integration, the reconstructed function $f(G)$ is expressed parametrically as $f(G(t))$ by employing the relation between the Gauss--Bonnet invariant and the cosmic time.

The numerical reconstruction is displayed in Fig.~\ref{fig:Fig2b}. Figure~\ref{fig:Fig2b} shows that the reconstructed $f(G)$ is characterized by a monotonically decreasing pattern with respect to $G$. At lower curvatures, the modified gravity corrections are more prominant and gradually decreases as the curvature increases. This indicates the significance of the Gauss--Bonnet modification as the spacetime curvature is weak. This influence is gradually decreased in higher-curvature regimes. Such behaviour allows the modified gravity sector to play a prominent role in driving the late-time cosmic acceleration without introducing large deviations from the standard gravitational dynamics in regimes of comparatively higher curvature.

\section{Cosmological Parameters}

In this section, we investigate the cosmological implications of the reconstructed model by analysing the evolution of the Hubble parameter, the effective equation of state parameter, and the deceleration parameter. The analysis is carried out using the reconstructed hybrid expansion law given by Eq.~\eqref{Eq:22}, along with the corresponding expressions for the Hubble parameter and its time derivative in Eqs.\eqref{Eq:23} and \eqref{Eq:24}. Their evolutionary pattern provides a description of the transition from the early decelerating phase to the present accelerated expansion. Thus, we can assess the cosmological viability of the reconstructed $f(G)$ gravity model. For the reconstructed functional form of $f(G)$, we now investigate the cosmological implications of the reconstructed modified gravity model. Because of the fact that it is possible to interpret the geometric corrections in $f(G)$ gravity as an effective cosmic fluid, the cosmological evolution can be conveniently described through the corresponding effective energy density and pressure. The effective equation of state parameter is defined as
\begin{equation}
w_{\rm eff}=\frac{p_{\rm eff}}{\rho_{\rm eff}},
\label{Eq:weff}
\end{equation}
where
\begin{align}
\rho_{\rm eff} &= \rho_{\rm ghost}+Gf_G-f(G)-24H^3\dot f_G, \nonumber\\
p_{\rm eff} &= p_{\rm ghost}+f(G)-Gf_G
+8H^2\ddot f_G
+16H\dot H\dot f_G
+16H^3\dot f_G.\nonumber
\end{align}

The reconstructed function $f(G)$ obtained in the previous section is substituted into the above expressions together with the hybrid expansion law. Hence, we have the reconstructed $w_{eff}$ as

\begin{equation}
\label{Eq:weff}
\begin{aligned}
w_{\rm eff}
&=\frac{p_{\rm ghost}+f(G)-Gf_G+8H^2\ddot{f}_G
+16H\dot{H}\dot{f}_G+16H^3\dot{f}_G}
{\rho_{\rm ghost}+Gf_G-f(G)-24H^3\dot{f}_G}.
\end{aligned}
\end{equation}
In the present viscous framework, the effective equation of state parameter can now be written as
\begin{equation}
w_{\rm eff} = \frac{w_{\rm ghost}(\alpha H+\beta H^2) -3H(\xi_0+\xi_1H) +f(G)-Gf_G +8H^2\ddot f_G +16H\dot H\dot f_G +16H^3\dot f_G}{\alpha H+\beta H^2 +Gf_G-f(G) -24H^3\dot f_G}.
\label{Eq:weff}
\end{equation}
In \eqref{Eq:weff}, $w_{ghost}$ is given by Eq.~\eqref{Eq:28} and it combines viscosity, ghost dark energy, and the reconstructed $f(G)$ into a single effective equation of state parameter. From Eq.~\eqref{Gt}, it is understandable that if $t \to \infty$, $G \to Constant$. Hence $f(G_{\infty}) \to Constant$, and $\dot{G} \to 0$. Then $\dot{f_G} \to 0$ and $\ddot{f_G} \to 0$. In this limiting scenario, we can have from \eqref{Eq:weff}
\begin{equation}
\lim_{t\to\infty}w_{\rm eff}
=\frac{w_{{\rm ghost},\infty}\left(\alpha\lambda+\beta\lambda^{2}\right)
-3\lambda\left(\xi_{0}+\xi_{1}\lambda\right)+f(G_{\infty})
-G_{\infty}f_{G}(G_{\infty})}{\alpha\lambda+\beta\lambda^{2}+G_{\infty}f_{G}(G_{\infty})
-f(G_{\infty})},
\label{Eq:weff_inf}
\end{equation}
From \eqref{Eq:weff_inf} we have in the limiting case an asymptotic de Sitter behaviour of the reconstructed effective equation of state parameter. Equation~\eqref{Eq:weff_inf} represents the asymptotic effective equation of state of the reconstructed viscous $f(G)$ gravity in the late-time limit $(t\rightarrow\infty)$. Consequently, the effective equation of state ceases to evolve dynamically and asymptotically approaches the constant value given by Eq.~\eqref{Eq:weff_inf}. This behaviour indicates that the reconstructed modified Gauss--Bonnet gravity naturally admits a stable late-time cosmological attractor.

We can have some physically important limiting cases from Eq.~\eqref{Eq:weff_inf}. In the absence of bulk viscosity we have $(\xi_{0}=\xi_{1}=0)$. In this case, the asymptotic cosmological evolution is governed by the the generalized ghost dark energy and the reconstructed $f(G)$ gravity. On the other hand, when the Gauss--Bonnet modification is not there, i.e. $f(G)=0$, we have Eq.~\eqref{Eq:weff_inf} getting reduced to
\begin{equation}
\lim_{t\rightarrow\infty}w_{\rm eff} =w_{{\rm ghost},\infty}-\frac{3(\xi_{0}+\xi_{1}\lambda)}{\alpha+\beta\lambda},
\end{equation}
which corresponds to the asymptotic equation of state of the viscous generalized ghost dark energy model within Einstein gravity. Furthermore, if we consider the scenario where the viscous and modified gravity contributions become negligible, the effective equation of state approaches the asymptotic ghost dark energy equation of state and the standard late-time cosmological behaviour is recovered.

\section{Classical Stability of the Reconstructed Model}

The classical stability of the reconstructed interacting viscous generalized QCD ghost dark energy model is examined through the squared sound speed, defined as
\begin{equation}
v_s^2=
\frac{\dot p_{\rm eff}}
{\dot\rho_{\rm eff}},
\label{Eq:vs}
\end{equation}
where the effective energy density and pressure are given by Eqs.~\eqref{8} and \eqref{9}. Differentiating these equations with respect to cosmic time, we obtain
\begin{align}
\dot{\rho}_{\rm eff}
&=
\dot{\rho}_{\rm ghost}
+G\dot f_G
-72H^2\dot H\,\dot f_G
-24H^3\ddot f_G,
\label{Eq:rhodot}
\\
\dot p_{\rm eff}
&=
\dot p_{\rm ghost}
-G\dot f_G
+8H^2\dddot f_G
+\left(32H\dot H+16H^3\right)\ddot f_G
\nonumber\\
&\qquad
+\left(16\dot H^{\,2}
+16H\ddot H
+48H^2\dot H\right)\dot f_G.
\label{Eq:pdot}
\end{align}

Consequently, using Eqs. \eqref{Eq:rhodot} and \eqref{Eq:pdot} the squared sound speed takes the form
\begin{equation}
v_s^2=
\frac{
\dot p_{\rm ghost}
-G\dot f_G
+8H^2\dddot f_G
+\left(32H\dot H+16H^3\right)\ddot f_G
+\left(16\dot H^{\,2}
+16H\ddot H
+48H^2\dot H\right)\dot f_G
}{
\dot\rho_{\rm ghost}
+G\dot f_G
-72H^2\dot H\,\dot f_G
-24H^3\ddot f_G
}.
\label{Eq:vs2}
\end{equation}

The quantities $\dot{\rho}_{\rm ghost}$ and $\dot{p}_{\rm ghost}$ are evaluated using the reconstructed ghost dark energy model described by Eqs.~\eqref{Eq:6} and \eqref{Eq:7}, while the functions $f(G)$, $f_G$, $\dot f_G$, $\ddot f_G$, and $\dddot f_G$ are obtained from the reconstructed modified Gauss--Bonnet gravity discussed in the previous section. The terms containing $\dot{p}_{\rm ghost}$ and $\dot{\rho}_{\rm ghost}$ arise from the interacting viscous ghost dark energy. On the other hand, the terms involving $\dot{f}_G$, $\ddot{f}_G$, and $\dddot{f}_G$ represent the contribution of the modified Gauss--Bonnet gravity. As these terms are multiplied by different powers of $H$ and its derivatives, their effect seems to bedependent on the stage of cosmic evolution. At early times, the curvature and the expansion rate are high. At this epoch,  the modified gravity terms contribute more significantly. As the universe reaches $\dot{H}\rightarrow0$, these contributions gradually become smaller, and the evolution is governed mainly by the asymptotic behaviour of the model.

\subsection{Cosmological Implications of a Reconstruction-Inspired \texorpdfstring{$f(G)$}{f(G)} Model}

The reconstruction procedure presented in the previous subsection has provided the numerical evolution of the modified functional form $f(G)$. In this present subsection, we adopt an analytical function motivated by the reconstructed behaviour.

The numerical reconstruction displayed in Fig.~\ref{fig:Fig2} shows that the reconstructed function remains smooth and monotonic throughout the cosmic evolution. Motivated by this behaviour, we consider the following reconstruction-inspired functional form
\begin{equation}
f(G)=\mu G^{n},
\label{Eq:fgmodel}
\end{equation}
where $\mu$ and $n$ are constant model parameters. This form is proposed as an effective phenomenological approximation for investigating the cosmological evolution and stability of the reconstructed model. We would like to mention that it may not be claimed that this power-law form mentioned above \eqref{Eq:fgmodel} as constituting a fully cosmologically viable $f(G)$ model in the sense of De Felice and Tsujikawa~\cite{DeFelice2009}. Rather, it serves as an analytical representation of the reconstructed behaviour. Moreover, for suitable choices of $n$, it reproduces the qualitative behaviour exhibited by the numerical reconstruction obtained in the previous section. At this juncture, it should be noted that the reconstruction-inspired power-law model given by Eq.~(56) is not intended to provide the unique analytical form of the numerically reconstructed function. Since the reconstruction is carried out numerically, different analytical functions may reproduce its qualitative behaviour over the relevant curvature range. The power-law form is adopted because of its mathematical simplicity, smoothness, and its ability to capture the monotonic behaviour of the reconstructed solution. Therefore, the parameters $\mu$ and $n$ should be regarded as effective phenomenological parameters that characterize the reconstructed behaviour rather than unique fundamental parameters of the underlying modified gravity theory.

For the model (\ref{Eq:fgmodel}), the first derivative is given by
\begin{equation}
f_G=\mu nG^{\,n-1},
\label{Eq:fgprime}
\end{equation}
while the successive time derivatives become
\begin{equation}
\dot f_G
=
\mu n(n-1)G^{\,n-2}\dot G,
\label{Eq:fgdot}
\end{equation}

\begin{equation}
\ddot f_G
=
\mu n(n-1)
\left[
(n-2)G^{\,n-3}\dot G^{\,2}
+
G^{\,n-2}\ddot G
\right],
\label{Eq:fgddot}
\end{equation}

and

\begin{equation}
\begin{aligned}
\dddot f_G
=&\,
\mu n(n-1)\Big[
(n-2)(n-3)G^{\,n-4}\dot G^{\,3}
\\
&
+3(n-2)G^{\,n-3}\dot G\,\ddot G
+G^{\,n-2}\dddot G
\Big].
\end{aligned}
\label{Eq:fgdddot}
\end{equation}

Substituting Eqs.~(\ref{Eq:fgmodel})$-$(\ref{Eq:fgdddot}) into Eqs.~(\ref{Eq:weff}) and (\ref{Eq:vs2}), the effective equation of state parameter and the squared sound speed can be expressed in terms of the cosmic time $t$. For $G$, we can utilize $H$, obtained using Eqs.~\eqref{Eq:22}$-$\eqref{Eq:24}. Consequently, the cosmological evolution of the reconstructed interacting viscous generalized QCD ghost dark energy model can be investigated analytically. At this juncture, we would like to mention that that since the Eq.~\eqref{Eq:fgmodel} is adopted as a reconstruction-inspired phenomenological approximation, the parameters $\mu$ and $n$ may be considered as effective fitting parameters rather than fundamental parameters of the underlying modified gravity theory. The parameter $\mu$ controls the overall magnitude of the Gauss--Bonnet contribution, whereas $n$ determines its nonlinear dependence on the Gauss--Bonnet invariant. 

\begin{figure*}[!t]
\centering

\begin{subfigure}[b]{0.32\textwidth}
    \centering
    \includegraphics[width=\linewidth]{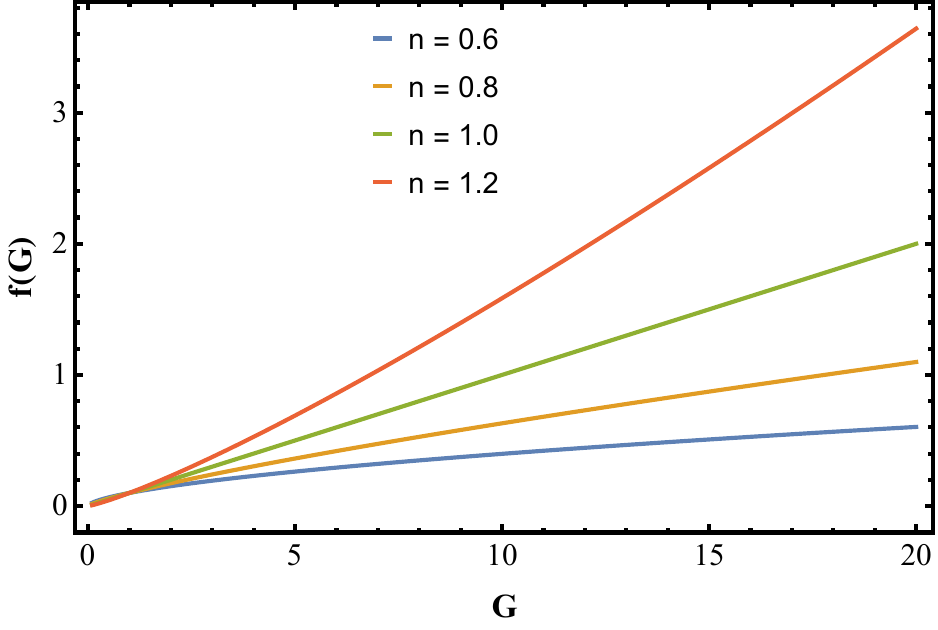}
    \caption{$f(G)$ as a function of $G$.}
    \label{fig:fG_power}
\end{subfigure}
\hfill
\begin{subfigure}[b]{0.32\textwidth}
    \centering
    \includegraphics[width=\linewidth]{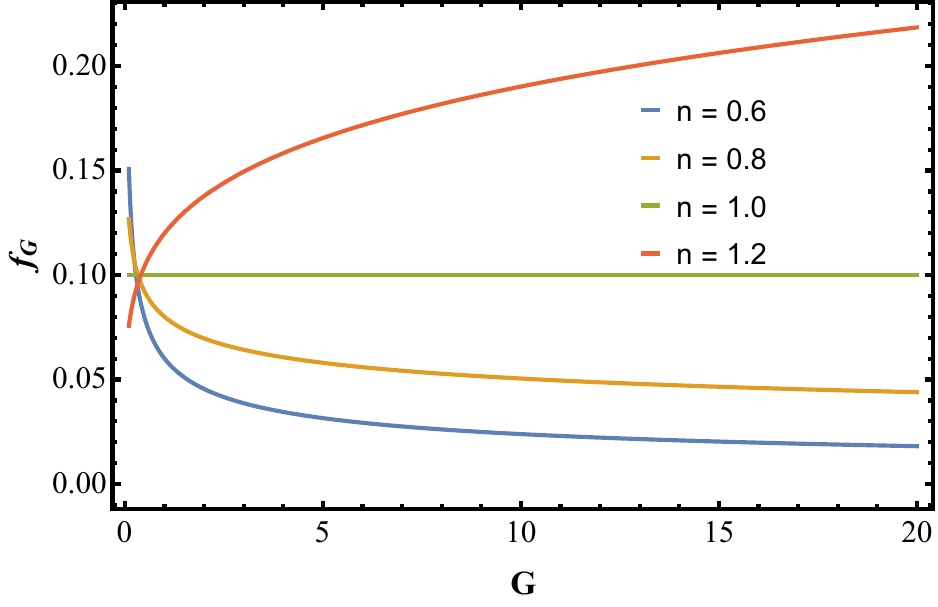}
    \caption{$f_G$ as a function of $G$.}
    \label{fig:fGprime_power}
\end{subfigure}
\hfill
\begin{subfigure}[b]{0.32\textwidth}
    \centering
    \includegraphics[width=\linewidth]{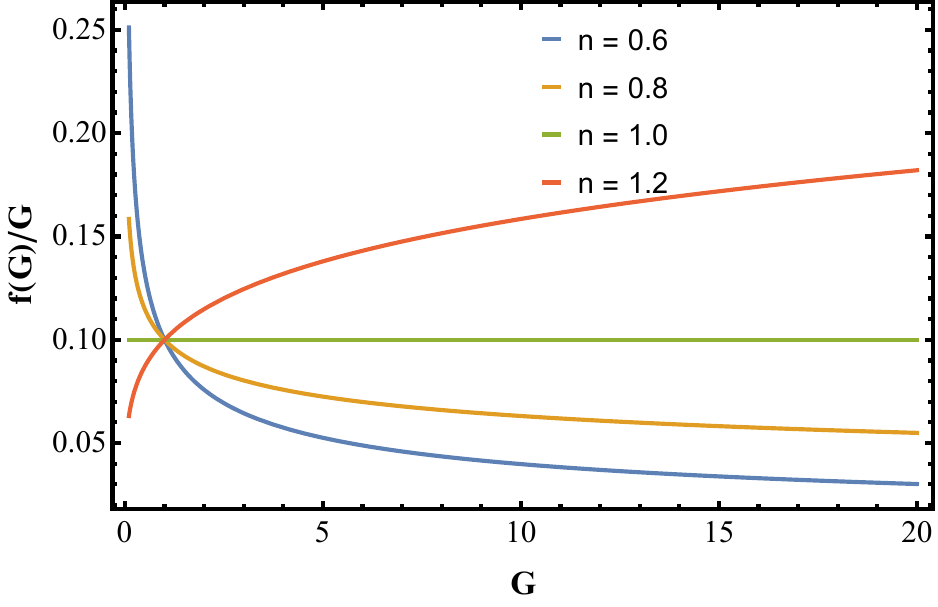}
    \caption{$f(G)/G$ as a function of $G$.}
    \label{fig:fGratio_power}
\end{subfigure}

\caption{Evolution of the reconstruction-inspired power-law Gauss--Bonnet model for different values of the power-law index $n$ with $\mu=0.1$. Panel (a) shows the reconstructed function $f(G)$, panel (b) illustrates its first derivative $f_G$, and panel (c) depicts the relative correction $f(G)/G$. The legends are shown inside each panel for clarity.}
\label{fig:powerlaw_all}
\end{figure*}

\begin{figure*}[!t]
\centering

\begin{subfigure}[b]{0.32\textwidth}
    \centering
    \includegraphics[width=\linewidth]{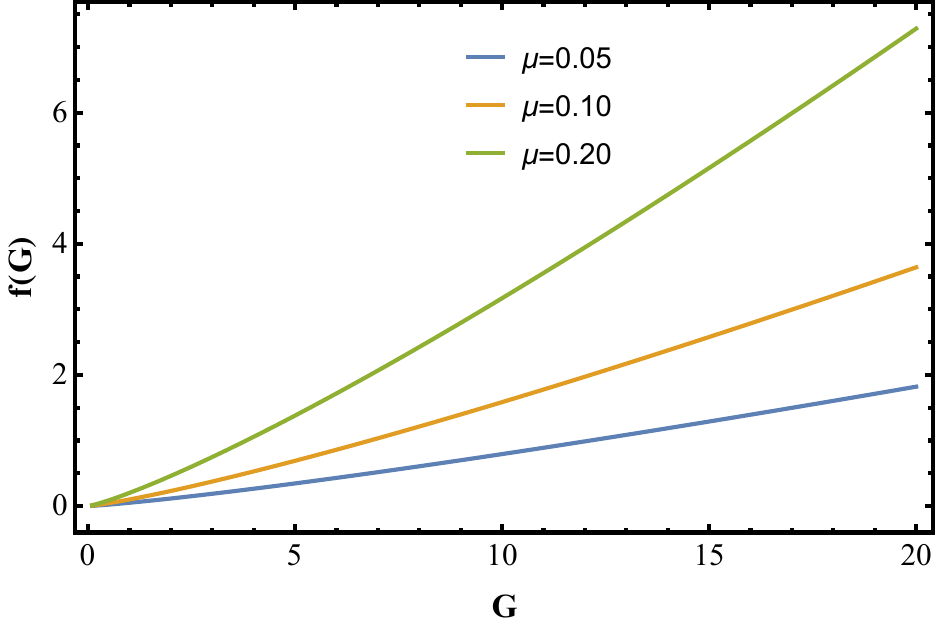}
    \caption{$f(G)$ as a function of $G$.}
    \label{fig:fG_power_mu}
\end{subfigure}
\hfill
\begin{subfigure}[b]{0.32\textwidth}
    \centering
    \includegraphics[width=\linewidth]{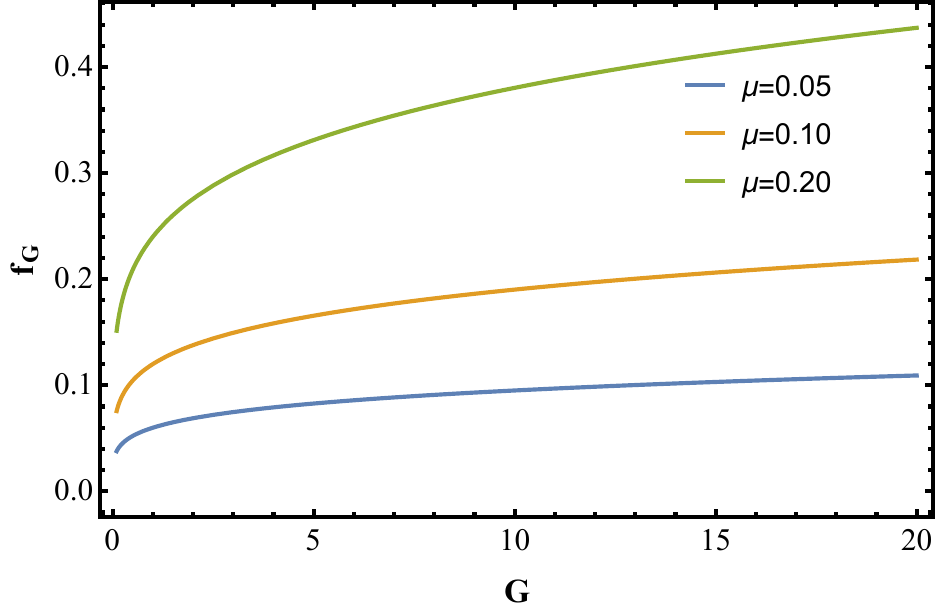}
    \caption{$f_G$ as a function of $G$.}
    \label{fig:fGprime_power_mu}
\end{subfigure}
\hfill
\begin{subfigure}[b]{0.32\textwidth}
    \centering
    \includegraphics[width=\linewidth]{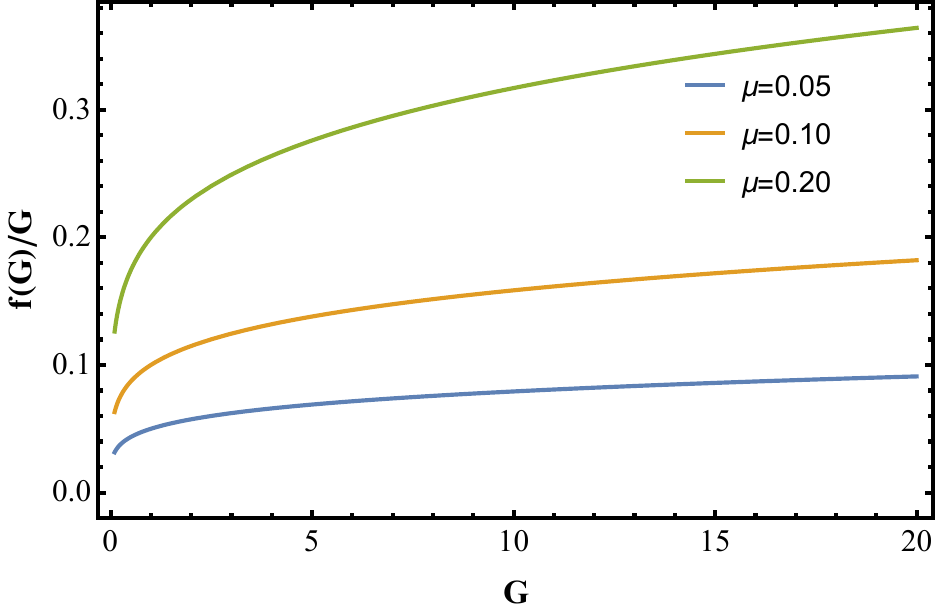}
    \caption{$f(G)/G$ as a function of $G$.}
    \label{fig:fGratio_power_mu}
\end{subfigure}

\caption{Evolution of the reconstruction-inspired power-law Gauss--Bonnet model for different values of $\mu$ with $n=1.2$. Panel (a) shows the reconstructed function $f(G)$, panel (b) illustrates its first derivative $f_G$, and panel (c) depicts the relative correction $f(G)/G$. The legends are shown inside each panel for clarity.}
\label{fig:powerlaw_all_mu}
\end{figure*}

Figures~\ref{fig:powerlaw_all} and \ref{fig:powerlaw_all_mu} illustrate the influence of the two effective parameters of the reconstruction inspired power-law model given by Eq.~\eqref{Eq:fgmodel}. Figure~\ref{fig:powerlaw_all} examines the effect of varying the power-law index $n$ while keeping the coupling parameter fixed, whereas Fig.~\ref{fig:powerlaw_all_mu} investigates the influence of the coupling parameter $\mu$ for a fixed value of $n$. From Fig.~\ref{fig:fG_power}, it is observed that the reconstructed function $f(G)$ increases monotonically with the Gauss--Bonnet invariant for all considered values of $n$. It is further noted that larger values of $n$ produce a steeper growth of $f(G)$. This indicates that the nonlinear dependence of $f(G)$ on the curvature invariant gets stronger with increase in $n$. The corresponding behaviour of the first derivative $f_G$, shown in Fig.~\ref{fig:fGprime_power}, demonstrates that the sensitivity of the modified gravity correction to the Gauss--Bonnet invariant is significantly affected by the choice of $n$. For $n<1$, $f_G$ gradually decreases with increasing $G$, whereas for $n>1$ it increases monotonically. The case $n=1$ represents the intermediate behaviour, corresponding to an almost constant derivative. Figure~\ref{fig:fGratio_power} further shows that the relative correction $f(G)/G$ decreases with increasing $G$ for $n<1$, remains nearly constant for $n=1$, and increases for $n>1$. These results indicate that the parameter $n$ primarily governs the nonlinear curvature dependence of the reconstructed Gauss--Bonnet correction. The effect of the parameter $\mu$ is presented in Fig.~\ref{fig:powerlaw_all_mu}. Unlike the power-law index, changing $\mu$ does not alter the qualitative behaviour of the reconstructed model. Instead, Figs.~\ref{fig:fG_power_mu}--\ref{fig:fGratio_power_mu} show that increasing $\mu$ uniformly increases the magnitudes of $f(G)$, $f_G$, and the relative correction $f(G)/G$ throughout the considered curvature range, while preserving their overall monotonic behaviour. Overall, we can interpret that $\mu$ primarily controls the overall strength of the modified Gauss--Bonnet contribution without changing its functional dependence on the curvature invariant.

\begin{figure*}[!ht]
\centering

\begin{subfigure}[b]{0.48\textwidth}
    \centering
    \includegraphics[width=\textwidth]{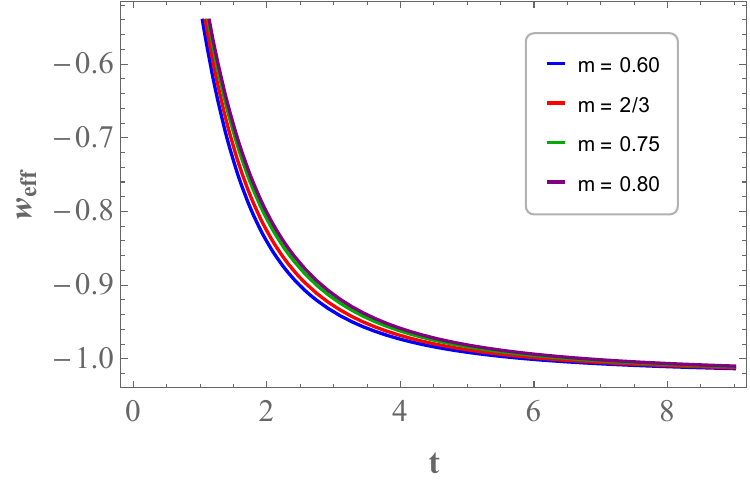}
    \caption{}
    \label{fig:weff_m}
\end{subfigure}
\hfill
\begin{subfigure}[b]{0.48\textwidth}
    \centering
    \includegraphics[width=\textwidth]{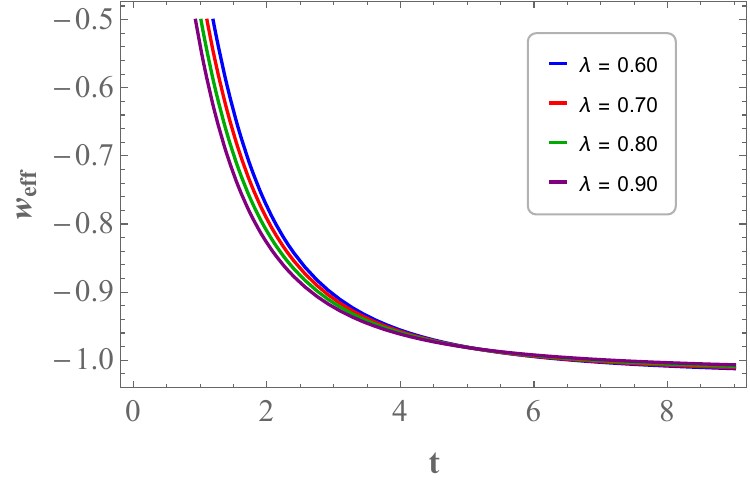}
    \caption{}
    \label{fig:weff_lambda}
\end{subfigure}

\caption{Time evolution of the effective equation of state parameter $w_{\rm eff}$ derived from Eqs.~\eqref{Eq:weff} and \eqref{Eq:fgmodel}. The left panel corresponds to different values of $m$, whereas the right panel illustrates the effect of varying $\lambda$, with the remaining parameters fixed at $n=1.2$, $\mu=0.10$, $\alpha=1.0$, $\beta=0.10$, $\xi_0=0.01$, $\xi_1=0.005$, and $b^2=0.03$.}
\label{fig:weff}
\end{figure*}

If Fig. \ref{fig:weff} we observe the time evolution of the effective equation of state parameter $w_{\rm eff}$ derived from Eqs.~\eqref{Eq:weff} and \eqref{Eq:fgmodel}  for the reconstruction-inspired power-law model $f(G)=\mu G^n$. The left panel depicts the influence of the parameter $m$ and the right panel demonstrates the effect of varying $\lambda$, while all remaining model parameters are kept fixed. In both panels, $w_{\rm eff}$ evolves from values around $-0.6$ at early epoch and gradually approaches the asymptotic value $w_{\rm eff}\rightarrow -1$, indicating a smooth transition from a quintessence-like phase to an effective de Sitter epoch. In a later stage of the universe, both the plots show that the $-1$ boundary is crossed and an asymptotic behaviour is found in the neighbourhood of $-1$ after crossing it. Hence, in a later stage, an asymptotic evolutionary behavior is observed, and a transition from quintessence to phantom is visible in a very late stage of the universe. However, it may be noted that in the phantom phase, the $w_{\rm eff}$ is asymptotic in the neighbourhood of $-1$. Therefore, the reconstructed interacting viscous generalized QCD ghost dark energy model naturally evolves toward a stable de Sitter phase while crossing phantom boundary but becoming asymptotic in the neighbourhood of $-1$, but just below it. 

\section{Thermodynamics}

In this section, we investigate the thermodynamic behaviour of the reconstructed interacting viscous generalized QCD ghost dark energy model. The analysis is performed on the apparent horizon, which is regarded as the physically relevant causal boundary of the FLRW universe. We adopt the standard Bekenstein--Hawking entropy associated with the apparent horizon \cite{Kruglov2025} and examine the validity of the generalized second law of thermodynamics (GSLT) \cite{Bekenstein1974,Izquierdo2006,Wu2008,Wall2009}.

For a spatially flat FLRW universe, the apparent horizon represents the natural causal boundary for studying the thermodynamic properties of the cosmic fluid \cite{Akbar2007,Cai2009,Sanchez2023}. Since the apparent horizon obeys the first law of thermodynamics (see \cite{Akbar2007,Cai2007UFL,Cai2009,Sheykhi2007,Sanchez2023,Kruglov2025}), it provides a consistent framework for examining the GSLT. The radius of the apparent horizon is given by $R_A=1/H$, where the Hubble parameter follows the reconstructed hybrid expansion law, $H=m/t+\lambda$. Consequently, the Hawking temperature associated with the apparent horizon is $T=1/(2\pi R_A)=H/(2\pi)$. Adopting the standard Bekenstein--Hawking entropy \cite{Bekenstein2008}, the horizon entropy is defined as $S_h=A/4$, where the horizon area is $A=4\pi R_A^2=4\pi/H^2$. Therefore, the horizon entropy takes the simple form
\begin{equation}
S_h=\frac{\pi}{H^2}.
\end{equation}
Differentiating with respect to cosmic time yields
\begin{equation}
\dot{S}_h=-\frac{2\pi\dot{H}}{H^3}.
\end{equation}
Using the reconstructed Hubble parameter, for which $\dot{H}=-m/t^2$, one finally obtains
\begin{equation}
\dot S_h = \frac{2\pi m}{t^2H^3}>0,
\label{Shdot}
\end{equation}
From the non-negative time derivative shown in Eq. \eqref{Shdot}, we understand that the entropy associated with the apparent horizon increases during the cosmological evolution.

The entropy of the cosmic fluid enclosed within the apparent horizon is obtained from the Gibbs equation \cite{Sharif2013,Tian2015},
\begin{equation}
TdS_f=d(\rho_{\rm eff}V)+p_{\rm eff}dV,
\label{gibbs}
\end{equation}
where the volume enclosed by the apparent horizon is
\begin{equation}
V=\frac{4\pi}{3H^3}.
\end{equation}

Differentiating \eqref{gibbs}, the above relation with respect to cosmic time gives
\begin{equation}
\dot S_f=\frac1T \left[ V\dot\rho_{\rm eff} + (\rho_{\rm eff}+p_{\rm eff}) \dot V\right].
\end{equation}

Since
\begin{equation}
T=\frac{H}{2\pi},
\end{equation}
the above expression becomes
\begin{equation}
\dot S_f
=
\frac{2\pi}{H}
\left[
V\dot\rho_{\rm eff}
+
(\rho_{\rm eff}+p_{\rm eff})
\dot V
\right].
\end{equation}

The generalized second law of thermodynamics is examined through the total entropy production,
\begin{equation}
\dot S_{\rm total} = \dot S_h+\dot S_f.
\end{equation}

Substituting the reconstruction-inspired power-law model
\begin{equation}
f(G)=\mu G^{n},\nonumber
\end{equation}
together with its derivatives, as already derived, into the effective energy density and pressure, we obtain
\begin{align}
\rho_{\rm eff}
&=\alpha H+\beta H^{2}+\mu(n-1)G^{n}-24\mu n(n-1)H^{3}G^{n-2}\dot G,
\label{rhoeffPL}
\\
p_{\rm eff}&=w_{\rm ghost}(\alpha H+\beta H^{2}) -3H(\xi_{0}+\xi_{1}H) +\mu(1-n)G^{n}
\nonumber\\
&
\quad
+8\mu n(n-1)H^{2} \left[(n-2)G^{n-3}\dot G^{2} +G^{n-2}\ddot G\right]
\nonumber\\
&
\quad
+16\mu n(n-1) (H\dot H+H^{3}) G^{n-2}\dot G .
\label{peffPL}
\end{align}

Using
\begin{equation}
V=\frac{4\pi}{3H^{3}},
\qquad
\dot V=-\frac{4\pi\dot H}{H^{4}},
\end{equation}
the Gibbs relation yields
\begin{equation}
\dot S_f
=\frac{8\pi^{2}}{3H^{4}}\dot\rho_{\rm eff}
-\frac{8\pi^{2}\dot H}{H^{5}}
(\rho_{\rm eff}+p_{\rm eff}).
\label{Sfdot1}
\end{equation}

Substituting Eqs.~(\ref{rhoeffPL}) and (\ref{peffPL}) into Eq.~(\ref{Sfdot1}), the fluid entropy production becomes
\begin{align}
\dot S_f
&=
\frac{8\pi^{2}}{3H^{4}}\frac{d}{dt}\Big[\alpha H+\beta H^{2}+\mu(n-1)G^{n}
\nonumber\\
&
\qquad\qquad
-24\mu n(n-1) H^{3}G^{n-2}\dot G
\Big]
\nonumber\\
&
-
\frac{8\pi^{2}\dot H}{H^{5}}
\Bigg[
(1+w_{\rm ghost})
(\alpha H+\beta H^{2})
-3H(\xi_{0}+\xi_{1}H)
\nonumber\\
&
\qquad\qquad
+8\mu n(n-1)H^{2}
\left(
(n-2)G^{n-3}\dot G^{2}
+
G^{n-2}\ddot G
\right)
\nonumber\\
&
\qquad\qquad
+16\mu n(n-1)
(H\dot H+H^{3})
G^{n-2}\dot G
\nonumber\\
&
\qquad\qquad
-24\mu n(n-1)
H^{3}G^{n-2}\dot G
\Bigg].
\label{Sfdotfinal}
\end{align}

For the reconstructed hybrid scale factor, we have already shown that 
\begin{equation}
H=\frac{m}{t}+\lambda,
\qquad
\dot H=-\frac{m}{t^{2}},
\qquad
\ddot H=\frac{2m}{t^{3}}, \nonumber
\end{equation}
the Gauss--Bonnet invariant is
\begin{equation}
G = 24H^{2}(H^{2}+\dot H),\nonumber
\end{equation}
so that
\begin{align}
\dot G &= 24\left[4H^{3}\dot H+2H\dot H(H^{2}+\dot H)+H^{2}\ddot H\right],
\\
\ddot G &=24\left[12H^{2}\dot H^{2}+4H^{3}\ddot H+2\dot H^{2}(H^{2}+\dot H)+2H\ddot H(H^{2}+\dot H)+4H\dot H\ddot H + H^{2}\dddot H \right],
\end{align}
where
\begin{equation}
\dddot H=-\frac{6m}{t^{4}}.
\end{equation}
Consequently, Eq.~(\ref{Sfdotfinal}) is completely determined by the cosmic time together with the model parameters
$(m,\lambda,\mu,n,\alpha,\beta,\xi_{0},\xi_{1},b^{2})$. Here the ghost dark energy equation-of-state parameter is obtained from Eq.~\eqref{Eq:25}. For numerical calculations, we adopt the explicit approximation by considering weak-interaction i.e. $b^{2}\ll1$
\begin{equation}
w_{\rm ghost}
\simeq
-1+
\frac{
9H^{2}(\xi_{0}+\xi_{1}H)
+\dfrac{m}{t^{2}}(\alpha+2\beta H)
}{
3H(\alpha H+\beta H^{2})
}.
\label{wghost}
\end{equation}

\begin{figure}[!ht]
\centering

\begin{subfigure}{0.48\textwidth}
    \centering
    \includegraphics[width=\linewidth]{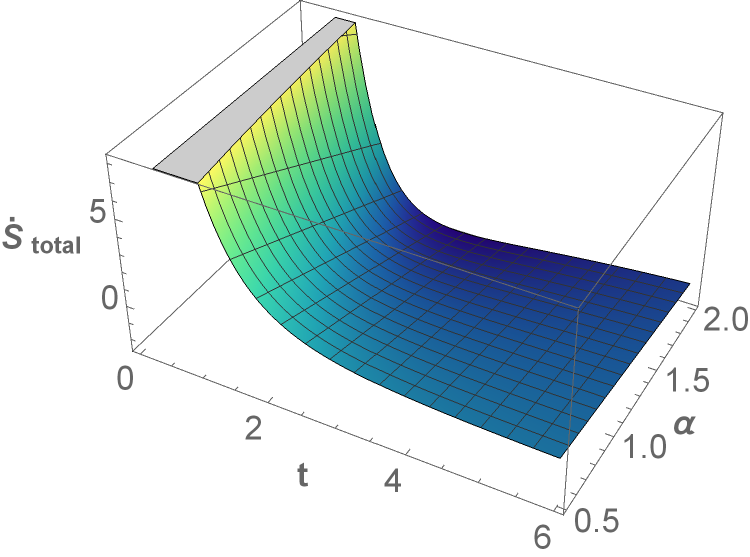}
    \caption{Evolution of $\dot{S}_{\rm total}$ with cosmic time $t$ and $\alpha$.}
    \label{fig:Sdot_alpha}
\end{subfigure}
\hfill
\begin{subfigure}{0.48\textwidth}
    \centering
    \includegraphics[width=\linewidth]{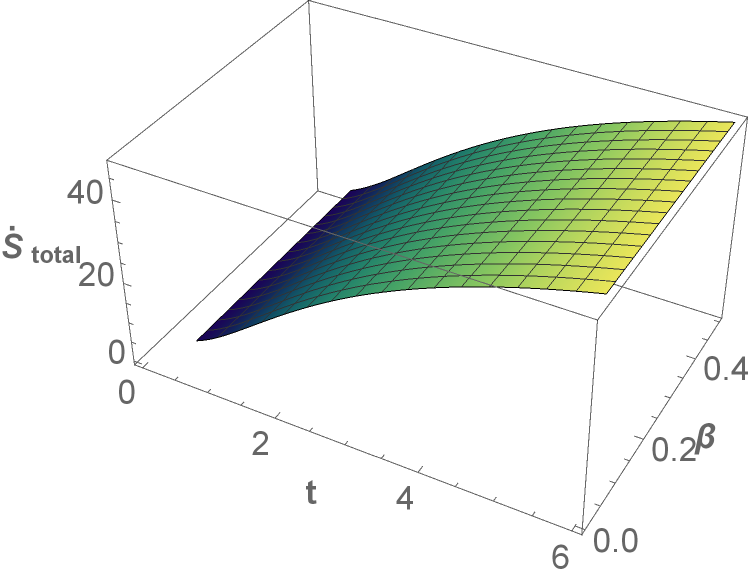}
    \caption{Evolution of $\dot{S}_{\rm total}$ with cosmic time $t$ and $\beta$.}
    \label{fig:Sdot_beta}
\end{subfigure}

\caption{Three-dimensional evolution of the total entropy rate $\dot{S}_{\rm total}$ for the reconstructed $f(G)$ model. The left panel illustrates the effect of varying the parameter $\alpha$, while the right panel shows the influence of varying $\beta$, with all other model parameters held fixed.}
\label{fig:Sdot_alpha_beta}
\end{figure}

The three-dimensional profiles of the total entropy rate $\dot{S}_{\rm total}$ presented in Fig.~\ref{fig:Sdot_alpha_beta} are obtained by using Eqs.~\eqref{Sfdotfinal}--\eqref{wghost}, where the total entropy rate is expressed as the sum of the horizon entropy rate and the fluid entropy rate, i.e., $\dot{S}_{\rm total}=\dot{S}_h+\dot{S}_f$. The horizon entropy rate and fluid entropy rate are computed from Eqs.~\eqref{Shdot} and \eqref{Sfdotfinal} respectively. The effective energy density and effective pressure entering Eq.~\eqref{Sfdotfinal} are evaluated from Eqs.\eqref{rhoeffPL} and \eqref{peffPL} respectively, together with the reconstruction-inspired power-law model $f(G)=\mu G^n$. The cosmological evolution is governed by the hybrid scale factor given in Eq.~\eqref{Eq:22}, with the corresponding Hubble parameter and Gauss--Bonnet invariant obtained from Eqs.~\eqref{Eq:23}--\eqref{Eq:24}. For the left panel, the parameter $\alpha$ is varied over the interval $0.5\leq\alpha\leq2$, while the remaining parameters are fixed at $m=2/3$, $\lambda=0.75$, $\beta=0.10$, $\mu=0.10$, $n=0.9$, $\xi_0=0.01$, and $\xi_1=0.005$. In the right panel, the parameter $\beta$ is varied in the interval $0.02\leq\beta\leq0.50$, and we take $\alpha=1$, $m=2/3$, $\lambda=0.75$, $\mu=0.10$, $n=0.9$, $\xi_0=0.01$, and $\xi_1=0.005$. The cosmic time is chosen within the interval $0.5\leq t\leq6$ to avoid the initial singularity. Figure~\ref{fig:Sdot_alpha_beta} depicts the three-dimensional evolution of the total entropy production rate $\dot{S}_{\rm total}$ for the reconstructed $f(G)$ model as functions of the cosmic time $t$ and the model parameters $\alpha$ and $\beta$. It is observed in both panels that $\dot{S}_{\rm total}$ remains strictly positive over the entire parameter space considered and hencet, it confirms that the reconstructed model is thermodynamically consistent. The left panel shows the dependence of $\dot{S}_{\rm total}$ on $\alpha$ and the right panel on $\beta$. It is observed that in the early phase, the rate of entropy production is higher than in the later dstage of the universe. Overall, it can be stated that the reconstructed $f(G)$ model satisfies GSLT for the chosen parameter values. Thus, the model remains thermodynamically viable and physically admissible over the entire evolutionary history considered.

\subsection{Thermodynamics with Barrow Entropy}
Recently, Barrow entropy has emerged as an important extension of the standard Bekenstein--Hawking entropy by incorporating possible quantum gravitational effects associated with the fractal deformation of the black-hole horizon. This generalized entropy has found wide applications in holographic dark energy and cosmological thermodynamics  \cite{Saridakis2020,Saridakis2021,BarrowBBN2021}. In particular, Barrow entropic dark energy has been shown to arise as a member of the generalized holographic dark energy family \cite{Nojiri_BarrowDE_2022}. Furthermore, Barrow entropy can be regarded as a particular realization of the more general Nojiri--Odintsov--Faraoni entropy framework, which provides a unified description of generalized entropy formalisms in gravitational and cosmological theories \cite{Nojiri_GeneralEntropy_2022,Nojiri_FaraoniEntropy_2022}. In this subsection, the thermodynamic behaviour of the reconstructed model is examined within the framework of Barrow entropy \cite{Barrow2021}, which incorporates possible quantum-gravitational corrections to the horizon geometry, and the validity of the generalized second law of thermodynamics is investigated following the formalism developed in Refs.~\cite{SaridakisJCAP2020,SaridakisPRD2020,SaridakisBasilakos2021}. In this framework, the entropy associated with the apparent horizon is modified from the standard Bekenstein--Hawking expression and is given by \cite{SaridakisBasilakos2021}

\begin{equation}
S_{B}=\left(\frac{A}{A_{0}}\right)^{1+\frac{\Delta}{2}},
\label{BarrowEntropy}
\end{equation}

where $A=4\pi R_{A}^{2}$ denotes the area of the apparent horizon, $A_{0}$ is the Planck area, and $\Delta$ $(0\leq\Delta\leq1)$ is the Barrow exponent which characterizes the quantum deformation of the horizon geometry. The standard Bekenstein--Hawking entropy is recovered in the limit $\Delta=0$.

For a spatially flat FLRW universe with the apparent horizon radius $R_{A}=\frac{1}{H}$ and $A=\frac{4\pi}{H^{2}}$, the Barrow entropy becomes
\begin{align}
S_B =\left(\frac{4\pi}{A_0}\right)^{1+\frac{\Delta}{2}}
H^{-(2+\Delta)}.
\end{align}
Differentiating with respect to the cosmic time gives
\begin{align}
\dot{S}_B
&=
\left(\frac{4\pi}{A_0}\right)^{1+\frac{\Delta}{2}}
\frac{d}{dt}
\left(H^{-(2+\Delta)}\right)\nonumber\\
&=
-(2+\Delta)
\left(\frac{4\pi}{A_0}\right)^{1+\frac{\Delta}{2}}
H^{-(3+\Delta)}
\dot{H},
\label{eq:SBdot}
\end{align}
Using the effective Friedmann equation, the Eq.~(\ref{eq:SBdot}) can be expressed in terms of the reconstructed effective fluid as
\begin{equation}
\dot{S}_{B} = \left(1+\frac{\Delta}{2}\right)
\left(\frac{4\pi}{A_{0}}\right)^{1+\frac{\Delta}{2}}
\left(\frac{\rho_{\rm eff}}{3}\right)^{-\frac{3+\Delta}{2}}
\left(\rho_{\rm eff}+p_{\rm eff}\right).
\label{SBdot_rho}
\end{equation}
The time derivative of total entropy is now obtained by combining the entropy variation of the apparent horizon, as evaluated using the Barrow entropy formalism, with the entropy production rate of the cosmic fluid derived from the Gibbs relation in Eq.~\eqref{Sfdotfinal}. The sign of $\dot{S}_{\rm total}$ is subsequently analyzed to examine the validity of the generalized second law of thermodynamics throughout the cosmic evolution.
\begin{figure}[htbp]
\centering
\includegraphics[width=0.7\textwidth]{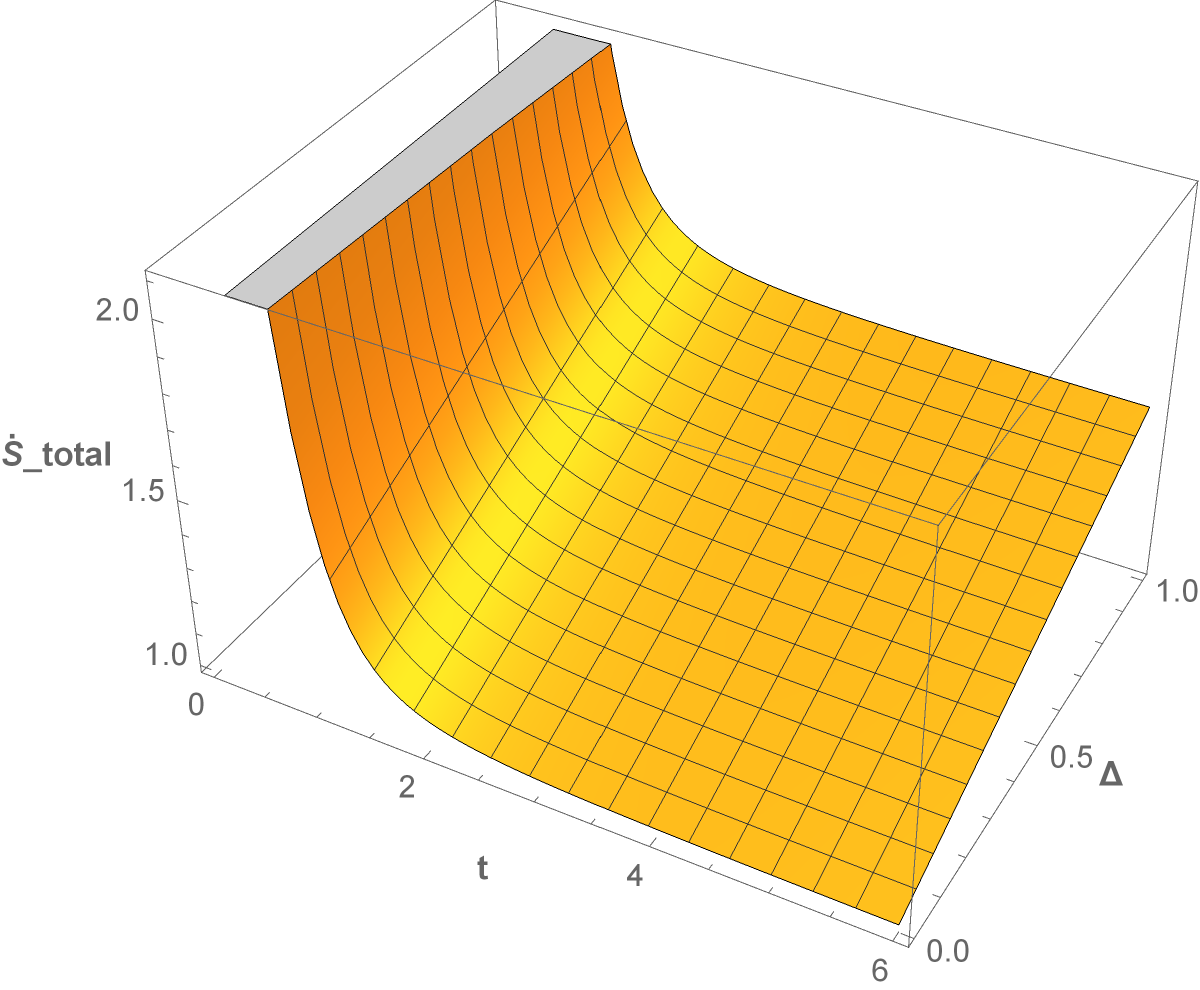}
\caption{Three-dimensional evolution of the total entropy rate $\dot{S}_{\rm total}$ as a function of the cosmic time $t$ and the Barrow exponent $\Delta$ for the reconstructed interacting viscous generalized QCD ghost dark energy model in $f(G)$ gravity. Sensitivity to 
the Barrow exponent $\Delta$ indicates the robustness of the thermodynamic behaviour against quantum geometrical corrections.}
\label{fig:SdotDelta}
\end{figure}

Figure~\ref{fig:SdotDelta} pictorially depicts the evolution of the total entropy rate $\dot{S}_{\rm total}$ as a function of the cosmic time $t$ and the Barrow exponent $\Delta$ for the reconstructed interacting viscous generalized QCD ghost dark energy model. The analysis is performed for $m=\frac{2}{3}$, $\lambda=0.75$, $\alpha=1.0$, $\beta=0.10$, $\mu=0.10$, $n=1.20$, $\xi_{0}=0.01$, and $\xi_{1}=0.005$, while $\Delta$ varies over the physically admissible interval $0\leq\Delta\leq1$. The Figure~\ref{fig:SdotDelta} shows that  $\dot{S}_{\rm total}$ remains strictly non-negarive throughout the cosmic evolution. This indicates the validity of the generalized second law of thermodynamics. From the Figure~\ref{fig:SdotDelta} it is also observed that the rate of entropy production has a monotone decreasing pattern with $t$ and from the early epochs of the cosmic evolution there is a sharp decay and subsequently becomes asymptotic at late times. Furthermore, the dependence of $\dot{S}_{\rm total}$ on $\Delta$ is also there. The $\dot{S}_{\rm total}$ appears to stay at higher level at higher values of $\Delta$. This indicates that the thermodynamic behaviour of the reconstructed model is mildly sensitive but stable against the quantum geometrical corrections implied by Barrow entropy. Thus, the interacting viscous generalized QCD ghost dark energy model reconstructed within the framework of $f(G)$ gravity can be considered as thermodynamically consistent over the evolutionary period under consideration. It may further be noted as the Barrow exponent $\Delta$ quantifies the quantum geometrical deformation of the apparent horizon, higher values of $\Delta$ imply stronger deviations from the standard Bekenstein--Hawking entropy. As a consequence, the Barrow entropy has marginally hgher contribution to the total entropy production and as a result there is  an increase in $\dot{S}_{\rm total}$. This moderate sensitivity  of the entropy production rate without violating the generalized second law of thermodynamics suggests that the inclusion of Barrow entropy reinforces the validity of the generalized second law of thermodynamics.

The present result is consistent with the general framework of Barrow entropy proposed in Ref.~\cite{SaridakisBasilakos2021}. While the validity of the generalized second law may depend on the cosmological background and the value of the Barrow exponent in the general analysis \cite{SaridakisBasilakos2021}, the reconstructed interacting viscous generalized QCD ghost dark energy model considered here satisfies the generalized second law throughout the cosmic evolution for the adopted parameter space.

\section{Comparison with Cosmic Chronometer Data}
\begin{figure}[!htbp]
\centering
\includegraphics[width=0.82\textwidth]{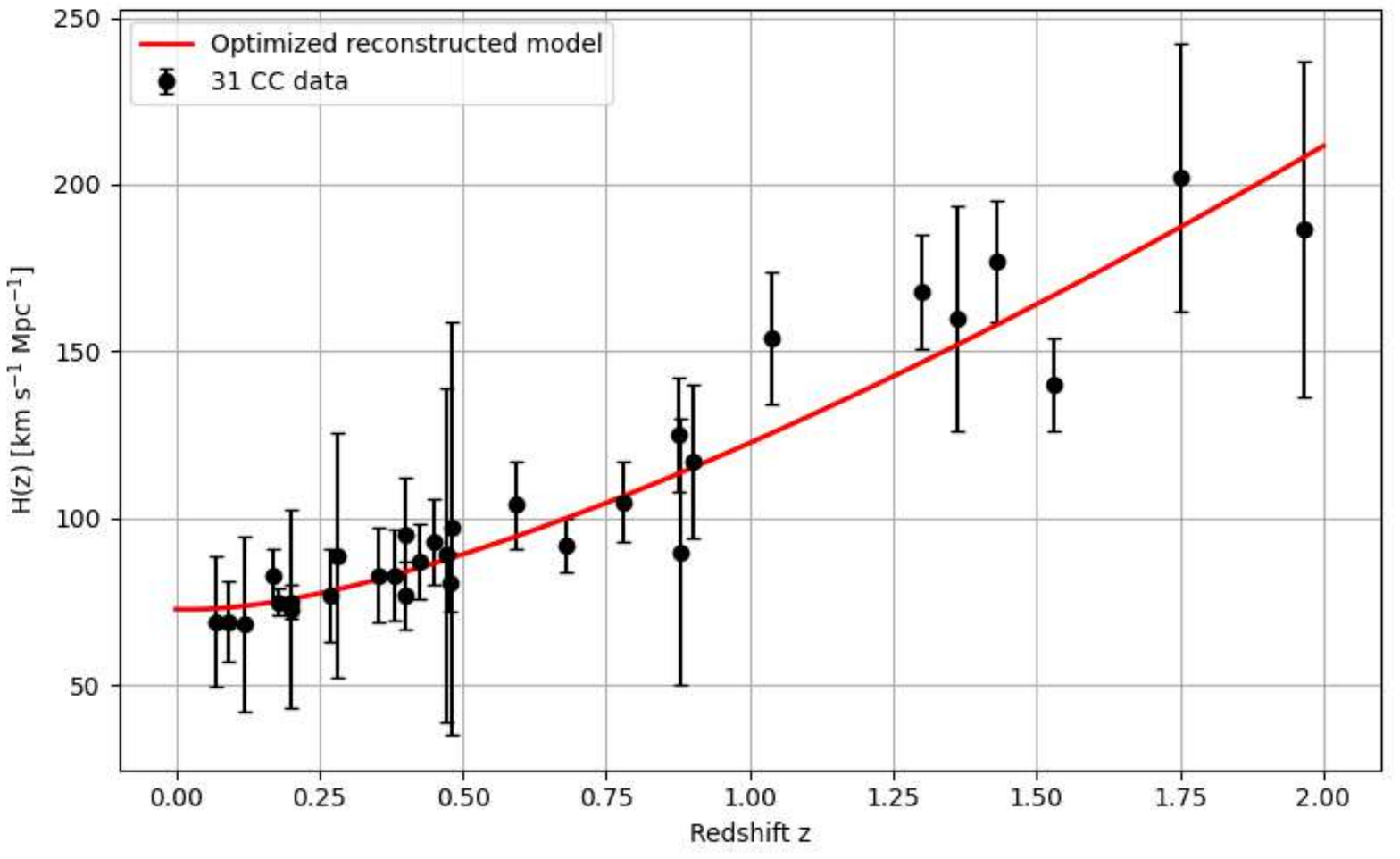}
\caption{Comparison of the reconstructed Hubble parameter with the 31 cosmic chronometer measurements. The solid red curve represents the optimized reconstructed interacting viscous generalized QCD ghost dark energy model in $f(G)$ gravity, while the black circles with error bars denote the observational cosmic chronometer data compiled by Yu \textit{et al.}~\cite{Yu2018}.}
\label{fig:CC}
\end{figure}

To examine the observational viability of the reconstructed interacting viscous generalized QCD ghost dark energy model, we compare the reconstructed Hubble parameter with the available 31 cosmic chronometer (CC) measurements, which provide direct determinations of the Hubble parameter over a wide redshift range without assuming a specific cosmological model \cite{Yu2018}. Since the modified Gauss--Bonnet contributions are incorporated into the effective fluid description, the reconstructed Hubble parameter is obtained from the effective Friedmann equation (Eq.~\eqref{Equation:6}) as
\begin{equation}
H_{\rm rec}(t)=\sqrt{\frac{\rho_m(t)+\rho_{\rm eff}(t)}{3}},
\end{equation}
where $\rho_m(t)$ is calculated from Eq.~\eqref{Eq:21}, and $\rho_{\rm eff}(t)$ is obtained from the power-law model $f(G)=\mu G^n$ using Eq.~\eqref{Equation:35}. Thus, the reconstructed Hubble parameter takes into account the effects of the modified $f(G)$ gravity through the effective fluid. The corresponding redshift dependence is obtained numerically from the reconstructed hybrid scale factor,
\begin{equation}
1+z=\frac{1}{t^{m}e^{\lambda t}},
\end{equation}
which enables the evaluation of $H_{\rm rec}(z)$ over the observed redshift range. The model parameters are then constrained by minimizing the standard chi-square statistic,
\begin{equation}
\chi^{2} =\sum_{i=1}^{31}\frac{\left[H_{\rm rec}(z_i)-H_{\rm obs}(z_i)\right]^{2}}{\sigma_i^{2}},
\end{equation}
where $H_{\rm obs}(z_i)$ and $\sigma_i$ denote the observed Hubble parameter and its corresponding uncertainty at the redshift $z_i$. The parameter values corresponding to the minimum $\chi^{2}$ are adopted for the comparison with the observational Hubble data. Figure~\ref{fig:CC} presents the resulting comparison between the reconstructed model and the 31 CC measurements.

Figure~\ref{fig:CC} compares the reconstructed Hubble parameter with the 31 cosmic chronometer measurements. It is observed that the reconstructed model follows the overall trend of the observational data over the entire redshift range. In spite of small deviations at a few redshifts, the reconstructed curve remains consistent with most of the observational measurements within their corresponding uncertainties. The fitting yields $\chi_{\rm min}^{2}=14.31$ and a reduced chi-square of $\chi_{\nu}^{2}=0.55$. Hence, it is a satisfactory agreement between the reconstructed model and the observational Hubble data. These results indicate the observational viability of the reconstruction model in the framework of $f(G)$ gravity and show that it is consistent with the observed cosmic expansion history.

\section{Conclusion}

In this work, we have investigated an interacting viscous generalized QCD ghost dark energy model in the framework of reconstructed $f(G)$ gravity. The cosmological dynamics of the model were first formulated by incorporating the interaction between dark matter and dark energy together with the effects of bulk viscosity. The corresponding hybrid expansion law was obtained through the reconstruction procedure using Eqs.~\eqref{Eq:27} and \eqref{Eq:28}, which naturally describes the transition from the matter-dominated era to the present accelerated expansion of the universe.

Using the reconstructed cosmological evolution, the equation of state parameter of the interacting viscous ghost dark energy was derived and analysed. The evolution shown in Fig.~\ref{Fig:figure1} indicates that the equation of state gradually approaches the $\Lambda$CDM limit at late times, showing that the model naturally evolves towards a dark-energy-dominated accelerating universe.

The modified Gauss--Bonnet function was reconstructed by solving the master reconstruction equation, Eq.~\eqref{master}. Since an exact analytical solution is difficult to obtain, the reconstruction was carried out numerically in both the early- and late-time regimes. The reconstructed function shown in Fig.~\ref{fig:Fig2} remains smooth and monotonic throughout the cosmic evolution, indicating that the reconstructed $f(G)$ model is numerically stable and physically acceptable. While asymptotic behaviors in modified Gauss--Bonnet gravity are often investigated separately in the early- and late-time limits, the work reported in the present paper deals with a unified framework based on a hybrid expansion law. The reconstruction of the functional form of $f(G)$ consistently describes both the high-curvature regime of the early universe ($G \sim t^{-4}$) and the asymptotic de Sitter phase at late times ($G \sim 24\lambda^{4}$),  and hence they connect the two cosmological epochs within a single model incorporating bulk viscous generalized QCD ghost dark energy and Barrow horizon thermodynamics.

Motivated by the numerical reconstruction, we introduced the reconstruction-inspired power-law model given by Eq.~\eqref{Eq:fgmodel}. The effective equation of state and the squared sound speed were then obtained from Eqs.~\eqref{Eq:weff} and \eqref{Eq:vs2}, respectively. Figures~\ref{fig:powerlaw_all} and \ref{fig:powerlaw_all_mu} illustrate the influence of the model parameters on the reconstructed function, while Fig.~\ref{fig:weff} shows that the effective equation of state evolves smoothly towards the de Sitter phase. These results suggest that the reconstructed model can successfully describe the late-time accelerated expansion of the universe.

The thermodynamic behaviour of the reconstructed model was also investigated. The horizon entropy rate was obtained from Eq.~\eqref{Shdot}, whereas the fluid entropy production rate was derived from Eq.~\eqref{Sfdotfinal} using the effective energy density and pressure given by Eqs.~\eqref{rhoeffPL} and \eqref{peffPL}. The total entropy production was analysed using the approximate ghost equation of state given by Eq.~\eqref{wghost}. Figure~\ref{fig:Sdot_alpha_beta} shows that the total entropy production rate remains positive throughout the considered parameter space, confirming the validity of the generalized second law of thermodynamics and the thermodynamic consistency of the reconstructed model. The study has also examined that thermodynamic consistency of the reconstructed interacting viscous generalized QCD ghost dark energy model within Barrow entropy framework (see Fig.~\ref{fig:SdotDelta}). Incorporating Barrow exponent $\Delta$, which corresponds to quantum geometrical correction, the study shows that the total entropy remains a non-decreasing function of $t$, thereby confirming the validity of the generalized second law of thermodynamics. Furthermore, a mild dependence of the entropy production rate has been observed on the Barrow exponent, indicating robustness against the quantum geometrical deformation of the horizon. These outcomes lead us to conclude that the inclusion of Barrow entropy does not affect the thermodynamic viability of the reconstructed $f(G)$ gravity model and hence it provides an additional support for its physical consistency to describe the late-time accelerated universe. Finally, in Fig.~\ref{fig:CC}, the reconstructed Hubble parameter is pictirially presented against observational CC data. The comparison with observational data exhibits good agreement with the 31 cosmic chronometer measurements. Hence,  Fig.~\ref{fig:CC} provides observational support for the proposed interacting viscous generalized QCD ghost dark energy model in $f(G)$ gravity. 

The reconstructed $f(G)$ function changes smoothly throughout the evolution of the universe, while the effective equation of state moves towards the de Sitter limit at late times. The stability analysis shows that the model is classically stable for suitable values of the model parameters.The thermodynamic analysis also confirms that the generalized second law of thermodynamics is satisfied. These results show that the interacting viscous generalized QCD ghost dark energy model in reconstructed $f(G)$ gravity can provide a consistent description of the late-time accelerated expansion of the universe. These results show that the interacting viscous generalized QCD ghost dark energy model in reconstructed $f(G)$ gravity provides a suitable description of the late-time accelerated expansion of the universe. The comparison with the 31 cosmic chronometer measurements further demonstrates that the reconstructed model is consistent with the observed cosmic expansion history. A more comprehensive observational study using additional datasets such as BAO, Pantheon+, SH0ES, and Planck will be considered in future work to further constrain the model parameters.

\section*{Acknowledgement}
The author Surajit Chattopadhyay acknowledges the visiting associateship of the Inter-University Centre for Astronomy and Astrophysics (IUCAA), Pune, India, where part of the work was carried out in January 2026.

\section*{Data Availability Statement}

No new experimental or observational data were generated or analysed in this study. All results were obtained from the theoretical framework and numerical calculations presented in the manuscript. The data supporting the findings of this study can be reproduced directly from the equations and model parameters provided in the article.

\end{document}